\documentclass[12pt]{article}
\usepackage{epsf}

\def\beq{\begin{equation}}
\def\eeq{\end{equation}}
\def\bea{\begin{eqnarray}}
\def\eea{\end{eqnarray}}

\def\nn{\nonumber}

\relax
\begin{document}

\begin{titlepage}
\begin{center}
{\large\bf Parafermionic theory with the symmetry $Z_N$, for $N$ even.}\\[.5in] 
{\bf Vladimir S.~Dotsenko\bf${}^{(1)}$,
 \bf Jesper Lykke Jacobsen\bf${}^{(2)}$
 \bf and Raoul Santachiara\bf${}^{(1)}$}\\[.2in]
{\bf (1)} {\it LPTHE\/}\footnote{Laboratoire associ\'e No. 280 au CNRS},
         {\it Universit{\'e} Pierre et Marie Curie, Paris VI\\
               Bo\^{\i}te 126, Tour 16, 1$^{\it er}$ {\'e}tage,
               4 place Jussieu, F-75252 Paris Cedex 05, France.}\\
{\bf (2)} {\it Laboratoire de Physique Th\'eorique et Mod\`eles
               Statistiques, \\
               Universit\'e Paris-Sud, B\^atiment 100, F-91405 Orsay,
               France.}\\[.2in]
dotsenko@lpthe.jussieu.fr, jacobsen@ipno.in2p3.fr,
santachiara@lpthe.jussieu.fr 
\end{center}

\underline{Abstract.}

Following our previous papers \cite{ref1,ref2} we complete the
construction of the parafermionic theory with the symmetry $Z_{N}$ based on
the second solution of Fateev-Zamolodchikov for the corresponding
parafermionic chiral algebra. In the present paper we construct the $Z_{N}$
parafermionic theory for $N$ even. Primary operators are classified according
to their transformation properties under the dihedral group
($Z_{N}\times Z_{2}$, where $Z_{2}$ stands for the $Z_{N}$ charge
conjugation), as two singlets, doublet $1,2,\ldots,N/2-1$, and
a disorder operator. In an assumed Coulomb gas scenario, the corresponding
vertex operators are accommodated by the Kac table based on the weight lattice
of the Lie algebra $D_{N/2}$. The unitary theories are representations of the
coset $SO_{n}(N)\times SO_{2}(N)/SO_{n+2}(N)$, with $n=1,2,\ldots$. We suggest
that physically they realise the series of multicritical points in statistical
systems having a $Z_{N}$ symmetry.

\end{titlepage}

\newpage

In this paper, we complete our previous work \cite{ref1,ref2} on the
construction of the $Z_{N}$ parafermionic conformal field theory based on the
second solution%
\footnote{In the following, we always refer to this {\em second solution}
of the parafermionic chiral algebra.}
of Fateev-Zamolodchikov for the corresponding parafermionic chiral algebra,
which has been given in the Appendix of Ref.~\cite{ref3}. The representation
theory for these $Z_{N}$ parafermions has a different structure for $N$ odd
and even. The theory with $N$ odd was the subject of Refs.~\cite{ref1,ref2}.
Its Kac table is based on the weight lattice of the algebra $B_{r}$
($N=2r+1$). The theory with $N$ even, which will be constructed in this paper,
has a Kac table based on the algebra $D_{r}$ ($N=2r$). Both cases, $N$ odd and
even, correspond to the coset \cite{ref4}:
\beq
\frac{SO_{n}(N)\times SO_{2}(N)}{SO_{n+2}(N)}.
\label{eq1}
\eeq
Here $SO_{n}(N)$ is the orthogonal group, with level $n$ for its affine
current algebra.

Physically, we expect that these new parafermionic theories should describe,
in particular, the multicritical points of statistical systems having a
$Z_{N}$ symmetry. For the moment this is only an expectation. To our
knowledge, the corresponding statistical models which would be realised on a
lattice, having $Z_{N}$ tricritical or multicritical fixed points, are not
known at present, except for the cases $Z_{2}$ and $Z_{3}$.

The local operator algebra of the parafermionic fields $\{\Psi^{k}(z)\}$ has
been given in the Appendix of Ref.~\cite{ref3}. With slightly different
notations, which we shall keep in this paper, it is reproduced in
Eqs.~(2.1)--(2.2) of Ref.~\cite{ref2}.

\begin{figure}
\begin{center}
 \leavevmode
 \epsfysize=250pt{\epsffile{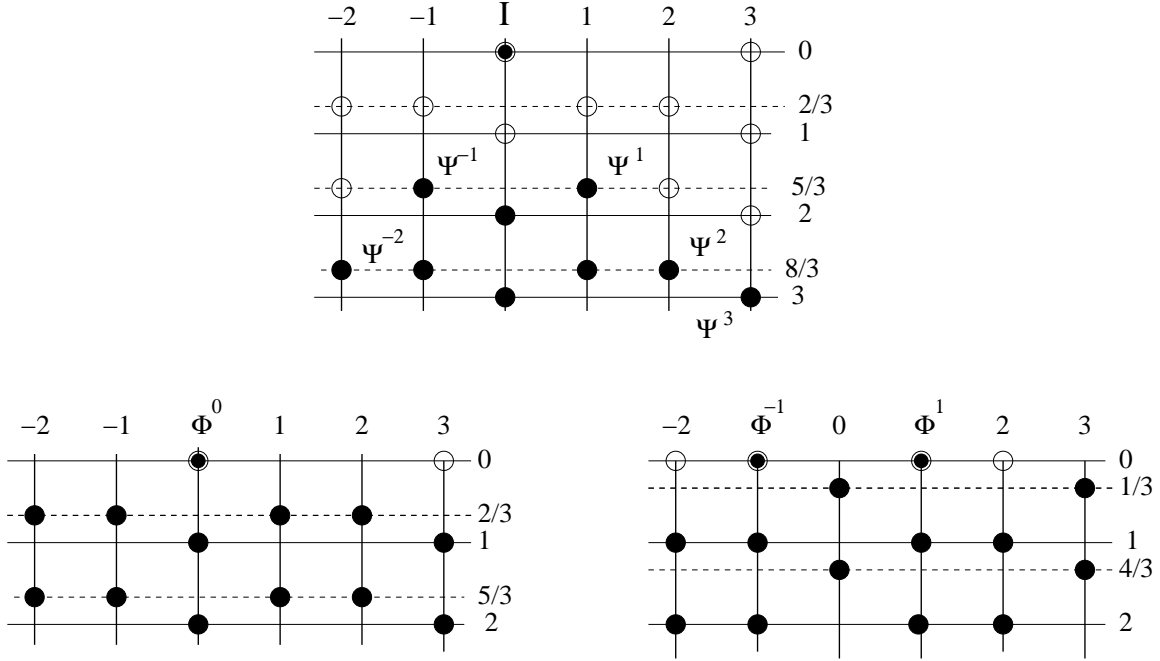}}
 \end{center}
 \protect\caption[3]{\label{fig1}Singlet and doublet modules of the $Z_6$
 theory as derived from the submodule structure of the identity operator
 module. Empty circles stand for empty states in the module.}
\end{figure}

Following Refs.~\cite{ref1,ref2}, the level structure of the representation
modules of singlet and doublet primary fields can be defined from the
conformal dimensions of the $\Psi^{k}(z)$ fields. Each of these dimensions
equals a level of the sector with $Z_N$ charge $k$, within the module of the
identity operator. Inside a fixed $Z_N$ sector, the Virasoro subalgebra acts,
making the level spacing equal to one. By taking submodules, the level
structure of all the modules (singlet, different doublets) can be defined (see
Ref.~\cite{ref2} for more details). In Fig.~\ref{fig1} we show as an example
the modules of the theory $Z_{6}$, which is the first non-trivial theory in
the family of $Z_{N}$ theories, for $N$ even.%
\footnote{The $Z_2$ theory is that of a free boson, $c=1$.
And the $Z_{4}$ theory
factorises into a direct product of two $N=1$ superconformal theories
(each of which having a $Z_2$ symmetry).}

We recall the formula for the conformal dimensions of the fields $\Psi^{k}(z)$
\cite{ref3}:
\beq
 \Delta_{k}=\Delta_{-k}=2\frac{k(N-k)}{N},
\label{eq2}
\eeq
where $k=1,2,\ldots,N/2$. For $Z_{6}$ one has $\Delta_{1}=5/3$,
$\Delta_{2}=8/3$, and $\Delta_{3}=3$, cf.~Fig.~\ref{fig1}.

A particular problem of defining the $Z_{N}$ theories with $N$ even appears
already at this point, when generalising the methods of Ref.~\cite{ref2} for
the case of $N$ odd. Considering the $Z_{6}$ theory as an example, we observe
that if we define the modules of $(\Phi^{-1}, \Phi^{1})$ and
$(\Phi^{-2},\Phi^{2})$ from the corresponding submodules within the module of
a singlet $\Phi^{0}$, as shown in Fig.~\ref{fig1}, these two doublet modules
will actually be identical. In other words, the doublet modules with $Z_N$
charge $q=\pm 1$ (``doublet 1'') and $q=\pm 2$ (``doublet 2'') are isomorphic
and thus only define one sector of the theory. In the same way, the module of
a singlet $\Phi^{3}$ is identical to that of the singlet $\Phi^{0}$.

This problem is general for all $Z_{N}$ with $N$ even. It can be traced back
to the fact that $\delta^{N/2-q}_{N/2-k}=\delta^{q}_{k}$, where
$\delta^{q}_{k}$ is the level spacing in the module of the doublet $q$,
cf.~Eq.~(\ref{eq6}) below. Thus, the primary fields $\{\Phi^{\pm q}\}$ define
distinct doublet modules only for $q=1,2,\ldots,q_{\rm max}$ with $q_{\rm
max}=\lfloor N/4 \rfloor$; here $\lfloor x \rfloor$ denotes the integer part
of $x$. For $N$ odd this problem was absent \cite{ref2} ($N/2$ in the above
formula is not an integer in this case), and accordingly one had
$q_{max}=\lfloor N/2 \rfloor$. Summarising, in the $Z_{N}$ theory with $N$
even, one half of the doublet primary fields appears to be missing.

We shall return to this problem of missing doublets later, to treat it
properly. For the moment we observe that in the case of the $Z_{6}$ theory one
would have a singlet, a doublet 1 and a disorder field.

The disorder fields are defined in a way similar to that described in
Refs.~\cite{ref1,ref2}, with modules having integer and half-integer levels.
We refer to the above-mentioned papers for details. The disorder sector does
not present major problems when generalising from $N$ odd to $N$ even. A
particular difference in the degeneracy structure of the disorder modules will
be mentioned later, in Appendix~\ref{sec_disorder}.

The decompositions of $\{\Psi^{k}(z)\}$ into mode operators, in the modules
of singlets and doublets, have the same form as in Eqs.~(2.12)--(2.14)
of Ref.~\cite{ref2}:
\bea
 \Psi^{k}(z)\Phi^{q}(0) &=& \sum_{n}\frac{1}{(z)^{\Delta_{k}-
 \delta^{q}_{k+q}+n}} A^{k}_{-\delta^{q}_{k+q}+n}\Phi^{q}(0), \label{eq3} \\
 A^{k}_{-\delta^{q}_{k+q}+n}\Phi^{q}(0) &=& 0,\quad \mbox{for }n>0,
 \label{eq4} \\
 A^{k}_{-\delta^{q}_{k+q}+n}\Phi^{q}(0) &=& 
 \frac{1}{2\pi i}\oint_{C_{0}}dz(z)^{\Delta_{k}-\delta^{q}_{k+q}+n-1}
 \Psi^{k}(z)\Phi^{q}(0),\quad \mbox{for }n\le 0. \label{eq5}
\eea
Here $\delta^{q}_{k}$ is the first descendent level (``gap'') in the module of
the doublet $q$, in the $Z_{N}$ charge sector $k$. It was established in
Ref.~\cite{ref2} that
\beq
 \delta^{q}_{k}=2\frac{(q^{2}-k^{2})}{N} \mbox{\ \ mod \ } 1.
\label{eq6}
\eeq
In the same way as in Ref.~\cite{ref2}, if $\delta^q_{k+q}$ in Eq.~(\ref{eq5})
happens to vanish, one can define the zero mode eigenvalues $\{h_{q}\}$:
\beq
 A^{\mp 2q}_{0}\Phi^{\pm q}(0)=h_{q}\Phi^{\mp q}(0).
\label{eq7}
\eeq
Note that the representations $\Phi^{q}$ are characterised by both
$\{h_{q}\}$ and the conformal dimension $\Delta_{q}$, the latter being
just the eigenvalue of the usual Virasoro zero mode $L_0$.

The commutation relations of the mode operators $\{A^{k}_{-\delta+n}\}$ have
been derived in Ref.~\cite{ref2}. In the present theory, with $N$ even, they
have the same general form and they are given by Eqs.~(2.17)--(2.24) of
Ref.~\cite{ref2}.

\begin{figure}
\begin{center}
 \leavevmode
 \epsfysize=300pt{\epsffile{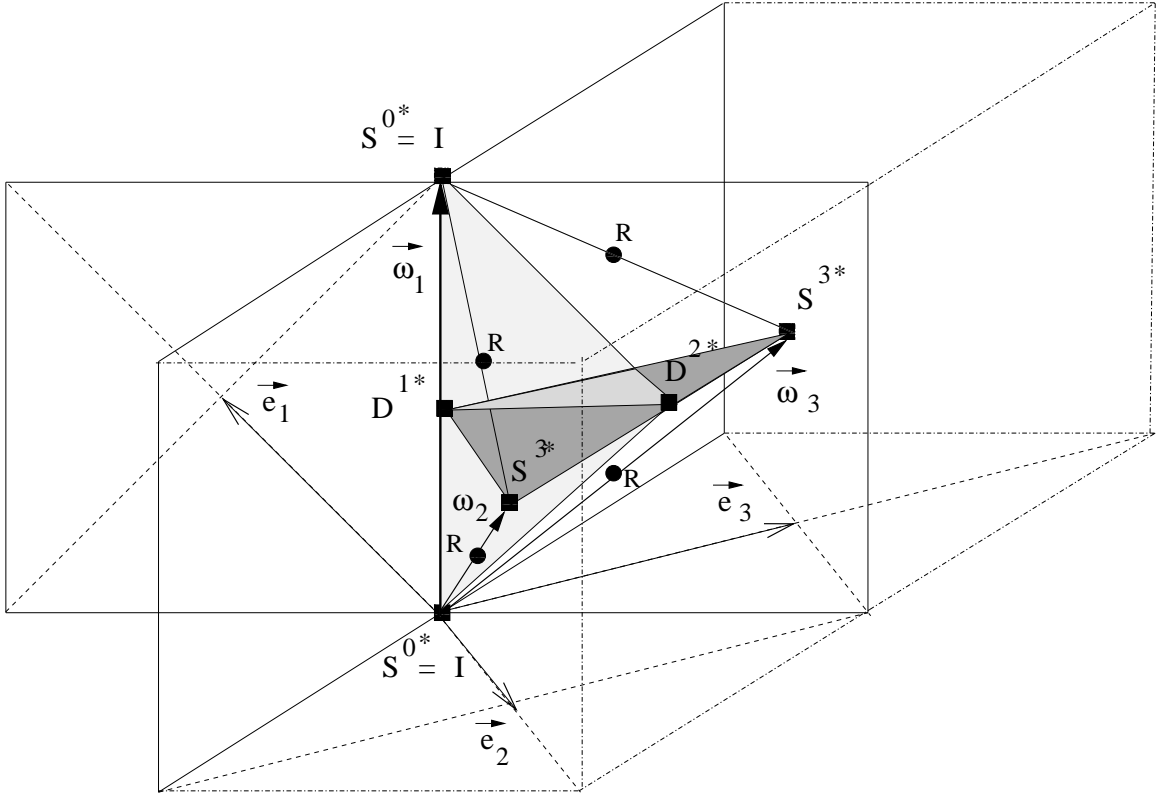}}
 \end{center}
 \protect\caption[3]{\label{fig2}Elementary cell of the theory
$Z_{6}$.  The fundamental operators, the independent ones after
applying the symmetry $Z_{2}\times Z_{2}$, are: the trivial singlet
$S^{0^{\ast}}=I=\Phi_{(1,1,1)}$, the doublets
$D^{1^{\ast}}=\Phi_{(1)}\equiv\Phi_{(2,1,1)}$, $D^{2^{\ast}}=
\Phi_{(2)}\equiv\Phi_{(1,2,2)}$, the non-trivial singlet
$S^{3^{\ast}}= \Phi_{(3)}\equiv\Phi_{(1,3,1)}$, and the disorder
operator $R=\Phi_{(1,2,1)}$.

The vectors $\vec{\beta}$ in Eq.~(\ref{eq9}), which constitute the lattice in
this figure, can be represented as
$\vec{\beta}=2\vec{\alpha_{0}}-\tilde{\vec{\beta}}$, with
$\tilde{\vec{\beta}}=\sum^{r}_{a=1}(\frac{1-n_{a}}{2}\alpha_{+}+\frac
{1-n'_{a}}{2}\alpha_{-})\vec{\omega}_{a}$. Accordingly, the vectors
corresponding to the sites of this lattice, with the origin at
$S^{0^{\ast}}=I$, represent the part
$-\tilde{\vec{\beta}}_{(1,1,1)(n'_{1},n'_{2},n'_{3})}=\sum^{3}_{a=1}\frac
{n'_{a}-1}{2}\alpha_{-}\vec{\omega}_{a}$ of
$\vec{\beta}_{(1,1,1)(n'_{1}, n'_{2},n'_{3})}$, while
$2\vec{\alpha}_{0}$ is being attributed to the origin.}
\end{figure}

\underline{The fundamental singlets, doublets and disorder operators},
one for each sector, are defined as operators whose modules are
degenerate at the first possible descendent levels. As already
mentioned, each primary operator of the theory is expected to occupy a
site of the $D_{r}$ weight lattice $(N=2r)$. Fundamental operators
occupy a portion of the lattice close to the origin, which we shall
refer to as the elementary cell. For $Z_{6}$, this elementary cell is
shown in Fig.~\ref{fig2}. In the figure, $\vec{\omega}_{1}$,
$\vec{\omega}_{2}$, $\vec{\omega}_{3}$ are the fundamental weights and
$\vec{e}_{1}$, $\vec{e}_{2}$, $\vec{e}_{3}$ are the simple root
vectors of the algebra $D_{3}$. We shall comment further on this
figure below.

The conformal dimensions of the operators are assumed to take the Coulomb gas
form
\bea
 \Delta_{\vec{\beta}}=\Delta^{(0)}_{\vec{\beta}}+B
 &=& \left( \vec{\beta}-\vec{\alpha}_{0} \right)^{2}
 -\vec{\alpha}_{0}^{2}+B, \label{eq8} \\
 \vec{\beta}\equiv\vec{\beta}_{(n_{1},n_{2},...n_{n})
 (n'_{1},n'_{2},...n'_{n})} &=&
 \sum^{r}_{a=1} \left( \frac{1+n_{a}}{2}\alpha_{+}+
 \frac{1+n'_{a}}{2}\alpha_{-} \right)
 \vec{\omega}_{a}, \label{eq9} \\
 \vec{\alpha}_{0} &=&
 \frac{(\alpha_{+}+\alpha_{-})}{2}\sum^{r}_{a=1}\vec{\omega}_{a},
 \label{eq10}
\eea 
where the Coulomb gas parameters $\alpha_{+}$, $\alpha_{-}$ are defined as
\beq
 \alpha_{+}=\sqrt{\frac{p+2}{2}},\quad \alpha_{-}=-\sqrt{\frac{p}{p+2}}
 \label{eq11}
\eeq
in accordance with the central charge expression for the coset (\ref{eq1})
\cite{ref3,ref4,ref2}:
\bea
 c &=& (N-1) \left( 1-\frac{N(N-2)}{p(p+2)} \right), \label{eq12} \\
 p &=& N-2+n. \label{eq13}
\eea
The constant $B$ in Eq.~(\ref{eq8}) is the {\em boundary term}, which takes, 
in general, different values for the different sectors of the theory.
To fully determine the Kac table of the theory, one needs to define these
sectors, work out the corresponding values of $B$, and assign the proper
sector label to each of the vectors $\vec{\beta}$.

Concerning the sector label assignment, we shall see that it is, in fact,
sufficient to properly place the fundamental operators in the elementary cell
of the Kac table. The assignment of sector labels to the rest of the lattice
will then be realised by Weyl-type reflections of the elementary cell. We
shall make this comment more precise in what follows.

First of all, the \underline{elementary cell of the $Z_{N}$ theory} is defined
as the finite part of the Kac table which corresponds, in particular, to the
physical domain of the trivial theory, i.e., the theory with level $n=0$ in
Eq.~(\ref{eq1}) and central charge $c=0$ in Eq.~(\ref{eq12}).

The unitary theories correspond to $n$ taking positive integer values in
Eqs.~(\ref{eq12})--(\ref{eq13}). For a particular unitary theory,
corresponding to a given value of the parameter $p$, the physical domain
of the Kac table is delimited as follows:
\bea
 n'_{1}+2\sum^{r-2}_{a=2}n'_{a}+n'_{r-1}+n'_{r} &\leq& p+1, \label{eq14} \\
 n_{1}+2\sum^{r-2}_{a=2}n_{a}+n_{r-1}+n_{r} &\leq& p-1, \label{eq15}
\eea
where $n'_{a}$, $n_{a}$ are positive integers. The reason is that on the
outer side of the hyperplanes
\bea
 n'_{1}+2\sum^{r-2}_{a=2}n'_{a}+n'_{r-1}+n'_{r} &=& p+2, \label{eq16} \\
 n_{1}+2\sum^{r-2}_{a=2}n_{a}+n_{r-1}+n_{r} &=& p \label{eq17}
\eea
the sites of the Kac table are occupied by ``ghosts'' (reflections of primary
submodule operators) which decouple, in correlation functions, from the
operators of the physical domain. For more comments, see
Refs.~\cite{ref1,ref2}. Note that the delimitations (\ref{eq14})--(\ref{eq15})
are different from those of the $WD_{r}$ conformal theory \cite{ref5}.

For the trivial $Z_{6}$ theory ($c=0$, $n=0$) the
inequalities (\ref{eq14})--(\ref{eq15}) take the form
\bea
 n'_{1}+n'_{2}+n'_{3} &\leq& 5, \label{eq18} \\
 n_{1}+n_{2}+n_{3}    &\leq& 3. \label{eq19}
\eea
The $\alpha_{+}$ side of the Kac table is reduced to one site,
$n_{1}=n_{2}=n_{3}=1$; a non-trivial domain exists only on the $\alpha_{-}$
side, cf.~Eq.~(\ref{eq18}). This domain is shown in Fig.~\ref{fig2}.

Also in the general case, the physical domain of the trivial $c=0$ theory is
confined by Eqs.~(\ref{eq14})--(\ref{eq15}) to trivial values $n_a=1$ of the
$\alpha_+$ indices. The $\alpha_-$ side is non-trivial and defines the
elementary cell whose operators we must identify. The identification of the
other operators in the Kac table will follow from this analysis.

We consider as being natural the following assumptions:
\begin{enumerate}
 \item The elementary cell, which, in particular, is the physical domain of the
 $c=0$ theory, should contain all possible types of operators:
 the singlet(s), all different doublets, and disorder. This implies,
 in particular, that the fundamental operators (the ones with degeneracies
 at the first available levels, one operator for each sector) should all be
 contained in the elementary cell of a given $Z_{N}$ theory.
 \item The theory $c=0$ being trivial, we shall assume that the conformal
 dimensions of the operators in the elementary cell all become equal to
 zero when $c=0$.
\end{enumerate}

The second assumption allows us to define the values of all the boundary
terms, i.e., the constant $B$ appearing in Eq.~(\ref{eq8}) for the dimensions
of primary operators. For a given $N=2r$, the $c=0$ theory corresponds to
$p=N-2$, cf.~Eqs.~(\ref{eq12})--(\ref{eq13}). {}From Eq.~(\ref{eq14}),
the elementary cell will be delimited by the inequality:
\beq
 n'_{1}+2\sum^{r-2}_{a=2}n'_{a}+n'_{r-1}+n'_{r}\leq N-1=2r-1.
 \label{eq20}
\eeq
The independent operators in this cell could be ordered in the following way:
\bea
\Phi_{(0)}  &\equiv& \Phi_{(1,1,1,\ldots,1,1)}=I\nn\\
\Phi_{(1)}  &\equiv& \Phi_{(2,1,1,\ldots,1,1)}\nn\\
\Phi_{(2)}  &\equiv& \Phi_{(1,2,1,\ldots,1,1)}\nn\\
\ldots      &\ldots& \ldots                   \nn\\
\Phi_{(r-2)}&\equiv& \Phi_{(1,1,1,\ldots,2,1,1)}\nn\\
\Phi_{(r-1)}&\equiv& \Phi_{(1,1,1,\ldots,1,2,2)}\nn\\
\Phi_{(r)}  &\equiv& \Phi_{(1,1,1,\ldots,1,3,1)}
\label{eq21}
\eea
plus one more operator
\beq
R=\Phi_{(1,1,1\ldots,1,2,1)}
\label{eq22}
\eeq
which will be shown to be a disorder operator. The operators in
Eq.~(\ref{eq21}) will be shown to be all possible singlets and doublets
admitted by the symmetry $Z_N$. We remind that, by the definition of the
elementary cell, all $\alpha_{+}$ indices have trivial values $n_{a}=1$,
and so we have suppressed them in Eqs.~(\ref{eq21})--(\ref{eq22}),
showing only the set of $\alpha_-$ indices $\{n'_{a}\}$.

All other operators inside the domain (\ref{eq20}) can be shown
to be related to the ones in Eqs.~(\ref{eq21})--(\ref{eq22}) 
by the symmetries of the elementary
cell. As an example, we show in Fig.~\ref{fig2} the elementary cell of
the theory $Z_{6}$, its operators, and its symmetries.

In general, the symmetry of the elementary cell of the theory $Z_{N}$, with
$N$ even, is of the type $Z_{2}\times Z_{2}$. It corresponds to reflections
into two mutually orthogonal hyperplanes, and will be made explicit below.
Eventually, one quarter of the
elementary cell contains all the independent operators. They are all the
fundamental ones, as we shall see shortly.

The geometry of the lattice generated by the vectors
$\{\frac{\vec{\omega}_{a}}{2},\ a=1,2,\ldots,r\}$ is contained in the matrix
of scalar products $\omega_{a,b}= (\vec{\omega}_{a},\vec{\omega}_{b})$. For
the $D_{r}$ lattice these scalar products take the following values:
\bea
 \omega_{a,b} &=& a,\quad a\leq b\leq r-2;\nn\\
 \omega_{a,r-1}=\omega_{a,r} &=& \frac{a}{2},\quad a\leq r-2;\nn\\
 \omega_{r,r}=\omega_{r-1,r-1} &=& \frac{r}{4};\nn\\
 \omega_{r-1,r} &=& \frac{r-2}{4}. \label{eq23}
\eea

It is appropriate at this point to make precise the normalisation conventions
which we assume in our formulae. We fix the overall normalisation of the
vectors $\{\vec{\omega}_{a}\}$ from the values of the scalar products
$\omega_{a,b}$ given in Eq.~(\ref{eq23}). The normalisation of the simple
roots $\{\vec{e}_{b}\}$ is then fixed by defining their scalar products with
$\{\vec{\omega}_{a}\}$ to be
\beq
 \left( \vec{\omega}_{a},\vec{e}_{b} \right) =
 \left| \vec{e}_{b} \right|^{2}\delta_{ab}.
\label{eq24}
\eeq
In the case of $D_{r}$ this implies that
\beq
|\vec{e}_{a}|^{2}=\frac{1}{2},\mbox{ for all $a$}.
\label{eq25}
\eeq
{}From the relation (\ref{eq24}) between $\{\vec{\omega}_{a}\}$ and
$\{\vec{e}_{a}\}$, one then finds that the decomposition of
$\vec{e}_{a}$ in the basis of $\{\vec{\omega}_{b}\}$ takes the form
\beq
 \vec{e}_{a}=\sum^{r}_{b=1}A_{ab}\frac{\vec{\omega}_{b}}{2},
\label{eq26}
\eeq
where $A_{ab}$ is the Cartan matrix
\beq
 A_{ab}=\frac{2(\vec{e}_{a},\vec{e}_{b})}{|\vec{e}_{b}|^{2}}.
\label{eq27}
\eeq

We shall return now to the problem of determining the boundary terms for all
the operators (\ref{eq21})--(\ref{eq22}) occupying the elementary cell. They
can be obtained from Eqs.~(\ref{eq8})--(\ref{eq11}) by demanding that the
conformal dimension of the corresponding operator vanish upon setting $p$
equal to $N-2=2r-2$, the value which corresponds to the $c=0$ theory,
cf.~Eqs.~(\ref{eq12})--(\ref{eq13}). Performing the calculations by using the
scalar products in Eq.~(\ref{eq23}), one finds that the constant $B$ should
take the following values for the operators $\Phi_{(a)}$ and $R$ defined in
Eqs.~(\ref{eq21})--(\ref{eq22}):
\bea
 B_{(a)} &=& \frac{a(r-a)}{4r},\quad a=0,1,2,\ldots,r \label{eq28} \\
 B_{R}   &=& \frac{r-1}{16}. \label{eq29}
\eea

One could object that in Eqs.~(\ref{eq8})--(\ref{eq11}) for the conformal
dimensions there is, in principle, a freedom in the overall normalisation of
the vectors $\{\vec{\omega}_{a}\}$, on which the values of the boundary terms
would depend, following the logic given above. In fact, this freedom is
removed by using the method of reflections which we shall describe now. The
normalisation will then get fixed exactly as in Eq.~(\ref{eq23}).

In a way analogous to the BRST structure of the (Virasoro algebra based)
minimal models \cite{ref6}, the reflections in the hyperplanes
which border the physical domain (\ref{eq14})--(\ref{eq15}) put in
correspondence the operators outside the physical domain with the degenerate
combinations of descendent fields inside the modules of physical operators
(i.e., operators positioned within the physical domain).

The \underline{simple reflections} $s_{\vec{e}_{a}} \equiv s_{a}$,
with $a=1,2,\ldots,r$, act on the weight lattice as the generators
of the Weyl group associated with $D_r$. They are defined by%
\footnote{Eq.~(\ref{eq30}) is taken from Eq.~(3.6) of Ref.~\cite{ref2},
but with the sign corrected.}
\beq
 s_{a}\vec{\beta}_{(1,1,\ldots,1)(n'_{1},n'_{2},\ldots,n'_{r})}
 = \vec{\beta}_{(1,1,\ldots,1)(n'_{1},n'_{2},\ldots,n'_{r})}-
 n'_{a}\alpha_{-}\vec{e}_{a}.
\label{eq30}
\eeq
Note again the trivial set of indices $\{n_{a}\}$ on the $\alpha_+$ side. 
As we shall continue to be interested in the operators of the elementary cell,
we shall suppress these indices again in the following. 
The vector $\vec{\beta}_{(\ldots)(n'_{1},n'_{2},\ldots,n'_{r})}$ in
Eq.~(\ref{eq30}) is to be taken outside the physical domain, in one of the
adjacent regions, while the result
of the reflection, i.e., the vector appearing on the right-hand side, should
belong to the physical domain.

In the case of unitary theories, the set of simple reflections (\ref{eq30})
has to be completed by a further reflection in the hyperplane (\ref{eq16}). We
denote this reflection by $s_{\vec{e}_{r+1}} \equiv s_{r+1}$. Expressed in
terms of the simple roots $\{\vec{e}_{a}\}$ (which, in a standard way,
correspond to screening operators) $s_{r+1}$ can be cast in a form similar to
that of Eq.~(\ref{eq30}):%
\footnote{Note the different signs of the second term on the
right-hand sides of Eqs.~(\ref{eq30})--(\ref{eq31a}). This means that
the simple reflections (\ref{eq30}) map ghosts {\em into} the physical
domain, while the reflection (\ref{eq31a}) is a mapping {\em out of}
the physical domain.}
\beq
 s_{r+1}\vec{\beta}_{(1,1,\ldots,1)(n'_{1},n'_{2},\ldots,n'_{r})}
 =\vec{\beta}_{(1,1,\ldots,1)(n'_{1},n'_{2},\ldots,n'_{r})}+n'_{r+1}\alpha_{-}
 \vec{e}_{r+1},
 \label{eq31a}
\eeq
with
\bea
 n'_{r+1} &=& N-n'_{1}-2\sum^{r-2}_{a=2}n'_{a}-n'_{r-1}-n'_{r}\nn\\
 \vec{e}_{r+1} &=& e_{1}+2\sum^{r-2}_{a=2}\vec{e}_{a}+\vec{e}_{r-1}+
 \vec{e}_{r},
\label{eq31}
\eea
where $\vec{e}_{r+1}$ is the affine simple root.

Since a given simple reflection connects a ghost operator (outside the
physical domain) and a degenerate (or singular) state inside the module of a
physical operator (inside the physical domain), the difference of conformal
dimensions of the ghost operator and the corresponding physical operator
should be compatible with the levels available in the module. These levels are
given by Eq.~(\ref{eq6}).

For the difference of dimensions one obtains, from Eq.~(\ref{eq8}),
\beq
 \Delta_{\vec{\beta}}-\Delta_{s_{a}\vec{\beta}}=\Delta^{(0)}_{\vec{\beta}}-
 \Delta^{(0)}_{s_{a}\vec{\beta}}+B_{\vec{\beta}}-B_{s_{a}\vec{\beta}}
 \label{eq32}
\eeq
Simple calculations using Eqs.~(\ref{eq8})--(\ref{eq11}), (\ref{eq24})
and (\ref{eq25})---as well as the definition of $\vec{\beta}$ and
$s_{a}\vec{\beta}$ in Eq.~(\ref{eq30})---lead to the result:
\beq
\Delta^{(0)}_{\vec{\beta}}-\Delta^{(0)}_{s_{a}\vec{\beta}}=-\frac{n'_{a}}{2}=
\frac{|n'_{a}|^{2}}{2}
\label{eq33}
\eeq
for $a=1,2,\ldots,r.$ 

It should be observed that the vector
\beq
 \vec{\beta}_{(\tilde{n}'_{1},\tilde{n}'_{2},...,\tilde{n}'_{r})}=
 \vec{\beta}_{(n'_{1},n'_{2},...,n'_{r})}-n'_{a}\alpha_{-}\vec{e}_{a}
\label{eq34}
\eeq
which must correspond to an operator inside the physical domain, satisfies
$\tilde{n}'_{a}=-n'_{a}$. This is due to the decomposition (\ref{eq26}) of
$\vec{e}_{a}$ in the basis of $\{\frac{\vec{\omega}_{b}} {2}\}$ and to the
fact that the diagonal elements of the matrix $A_{ab}$ are always equal to 2.
Now, by definition of the physical domain, the indices
$\tilde{n}'_{1},\tilde{n}'_{2},...,\tilde{n}'_{r}$ should all be positive. And
so, the index $n'_{a}$ of the ghost operator
$\vec{\beta}_{(n'_{1},n'_{2},...,n'_{r})}$ has to be negative. This explains
the last equality in (\ref{eq33}). This implies also, in the case of the
simple reflection $s_{a}$, that $|n'_{a}|$ is the index of the corresponding
physical operator, $\tilde{n}'_{a}=|n'_{a}|$.

Combining Eqs.~(\ref{eq32})--(\ref{eq33}) one finally obtains:
\beq
 \Delta_{\vec{\beta}}-\Delta_{s_{a}\vec{\beta}}=
 \frac{|n'_{a}|}{2}+B_{\vec{\beta}}-B_{s_{a}\vec{\beta}}
 \label{eq35}
\eeq
The boundary terms here carry indices $\vec{\beta}$ and $s_{a}\vec{\beta}$
in order to recall that these constants (which are independent of $c$, and
of the parameters $\alpha_{+}$, $\alpha_{-}$) take, in general, different
values for different sites of the $\vec{\beta}$ lattice, according to the
positioning of the various singlet, doublet and disorder sector operators.

Our purpose is to classify the operators (\ref{eq21})--(\ref{eq22}) of the
elementary cell as singlet, doublet and disorder operators, in accordance with
the $Z_{N}$ symmetry of the theory. For each of these operators, the position
on the weight lattice is known, so we know the first term on the right-hand
side of Eq.~(\ref{eq35}). The boundary terms of the operators
(\ref{eq21})--(\ref{eq22}) are known from Eqs.~(\ref{eq28})--(\ref{eq29});
this provides the third term on the right-hand side of Eq.~(\ref{eq35}),
$B_{s_{a}\vec{\beta}}$. The ghost boundary term $B_{\vec{\beta}}$ is not given
in advance. However, its value must be found among those of
Eqs.~(\ref{eq28})--(\ref{eq29}), since by Assumption 1 [made just after
Eq.~(\ref{eq19})] the operators of the elementary cell cover all possible
sectors of the theory.

On the other hand, as has been said before, the left-hand side of
Eq.~(\ref{eq35}) should match with the levels available in the modules. These
are given by the numbers $\delta^{q}_{k}$ in Eq.~(\ref{eq6}). This constraint
is actually even stronger, because a particular level in the module
corresponds only to a limited number of $Z_N$ charge sectors.
Thus, when computing the right-hand side of Eq.~(\ref{eq35})---trying
successively all possible values (\ref{eq28})--(\ref{eq29}) of the unknown
term $B_{\vec{\beta}}$---one needs to stay compatible with the available
levels, and at the same time the charge sector of the available state in the
module should match with the sector of the ghost. These constraints are
sufficiently strong to allow to identify almost uniquely the nature of all the
operators in the elementary cell.

This method has already been used in Ref.~\cite{ref2} for the case of odd $N$.
It leads to a unique identification of operators for $N$ sufficiently small,
namely $N\leq 13$. For larger $N$ extra solutions appear, which are however
non-regular in $N$. In the following, when applying this method to the case of
$N$ even, we shall ignore these sporadic possibilities and assume that we are
looking for a theory which is regular in $N$.

In the case of the theory with $N$ even there is an extra problem to
resolve, that of missing doublets. We have already mentioned that by
inspecting the structure of the modules, with the levels and $Z_{N}$
charge sectors defined by Eq.~(\ref{eq6}), one finds that the number
of different modules is only one half the number of expected representations
for the group $Z_{N}$. For instance, for $Z_{6}$ the modules of $q=0$
(singlet) and $q=\pm 1$ (doublet 1) are found to be distinct, while
the module of $q=\pm 2$ is identical to that of $q=\pm 1$ and $q=3$ is
identical to that of $q=0$.

Another mismatch with the number of representations of $Z_{N}$ is in the
number of different values of boundary terms. The values of the boundary
terms for the operators (\ref{eq21}) are given by Eq.~(\ref{eq28}), which
possesses a symmetry with respect to $a\rightarrow r-a\equiv N/2-a$. As a
result, the number of different values is again reduced by a factor of two. On
the other hand, the number of operators in Eq.~(\ref{eq21}) remains
identical to the number of different representations of $Z_{N}$.

To judge from these observations, one could be tempted to identify the
sectors of those operators in Eq.~(\ref{eq21}) which have identical
boundary terms, reducing in this way the number of sectors in the
theory by a factor of two.  This might appear to be consistent with
the number of different modules, which is also twice smaller.

This reduction could be obtained by assuming a different symmetry of
the elementary cell, reducing in this way the number of independent
operators in it. This new symmetry would be that of $Z_{4}$, replacing
the $Z_{2}\times Z_{2}$ symmetry which was assumed earlier. In the
elementary cell of the $Z_6$ theory, shown in Fig.~\ref{fig2}, this
$Z_4$ symmetry would be generated by a rotation of the cell through an
angle $\frac {\pi}{2}$ around the axis joining the sites
$D^{1^{\ast}}$ and $D^{2^{\ast}}$, followed by a reflection in a 
plane which is perpendicular to this axis and passes through its mid-point.

One could compare the symmetries of the elementary cell to those appearing in
the $W$-type theories based on the same Lie algebra. In Ref.~\cite{ref2}, the
symmetries of the elementary cells (and, more generally, of the physical
domains for the $c>0$ unitary theories) of the $Z_N$ theory with $N$ odd
was found to coincide with those of the $WB_r$ theory \cite{ref5}. For
even $N$, one could compare to the $WD_r$ theory \cite{ref5}, whose
elementary cell has a $Z_2 \times Z_2$ symmetry when $r$ is even,
and a $Z_4$ symmetry when $r$ is odd. In the particular case of the
$Z_6$ theory, the symmetry of the elementary cell would be $Z_4$ if
the correspondence with the $WD_3$ theory were to be maintained.

The only mismatch of the promising scenario suggested above is that the number
of different sectors of the theory would not correspond to the number of
representations of $Z_{N}$; it would be twice smaller. As a consequence, we
shall abandon the scenario described above and look for another one which
would respect the number of representations of $Z_{N}$.%
\footnote{At this point the knowledge of the actual symmetry of the theory is
crucial. The reference to the corresponding coset would not be
useful in any way.}

So far our definition of the modules has been based on the submodule structure
of the identity operator module, which is filled with the parafermionic
currents and their derivatives (and normal ordered products thereof)
\cite{ref1,ref2}. In particular, in this construction the $Z_{N}$ charge
sectors of the representation fields $\Phi^{q}$ were the same as those
of the parafermionic operators $\Psi^{k}(z)$, i.e., $q$ and $k$ were taking
the same set of values
\beq
 q,k=\pm 1,\pm 2,\ldots,\pm \lfloor N/2 \rfloor.
\label{eq36}
\eeq
The problem that has occurred in the case of the theory with $N$ even is that
this method only produces distinct modules for $|q| \le \lfloor N/4 \rfloor$.
Namely, the modules corresponding to $\Phi^{\pm q}$ with $\lfloor N/4 \rfloor
< q \leq \lfloor N/2 \rfloor$ are in fact identical to those of $0 \leq q
\leq \lfloor N/4 \rfloor$.

The attempt of recovering a number of sectors equal to the number of
representations of $Z_{N}$ has led us to suggest that one must also
consider half-integer values of $q$. Thus, we have assumed that the
allowed values of $q$ for the {\em primary}%
\footnote{I.e., primary with respect to the parafermionic algebra.}
representation fields should be
\beq
 q=0,\pm\frac{1}{2},\pm 1,\pm\frac{3}{2},\ldots,\pm \lfloor N/4 \rfloor.
\label{eq37}
\eeq
The levels in the modules corresponding to half-integer values of $q$ can
still be calculated from Eq.~(\ref{eq6}). In fact, this formula could alternatively be
re-derived by considering the associativity requirements for the four-point
function
\beq
 \left\langle \Phi^{-q}(\infty)\Psi^{-k}(1)\Psi^{k}(z)\Phi^{q}(0)
 \right \rangle = \frac{P_{l}(z)}
 {(1-z)^{2\Delta_{k}}(z)^{\Delta_{k}-\delta^{q}_{k+q}}}.
\label{eq38}
\eeq
Here $P_{l}(z)$ is a polynomial of degree $l$; the integer $l$ is determined
by the condition that $\Phi^{q}(0)$ be a primary (highest weight) field with
respect to the currents $\Psi^{k}(z)$, i.e., that the gaps $\delta^{q}_{k}$
have to take the smallest possible non-negative values.
It is sufficient to analyse the limit $z\to\infty$ of the function above,
by comparing the analytic form on the right-hand side of Eq.~(\ref{eq38})
with the expression obtained by using the operator algebra for the
fields on the left-hand side.

Within this approach one could perfectly well consider $q$ as being
half-integer valued, while the charges $k$ of the parafermions remain integer
valued.

\begin{figure}
\begin{center}
 \leavevmode
 \epsfysize=90pt{\epsffile{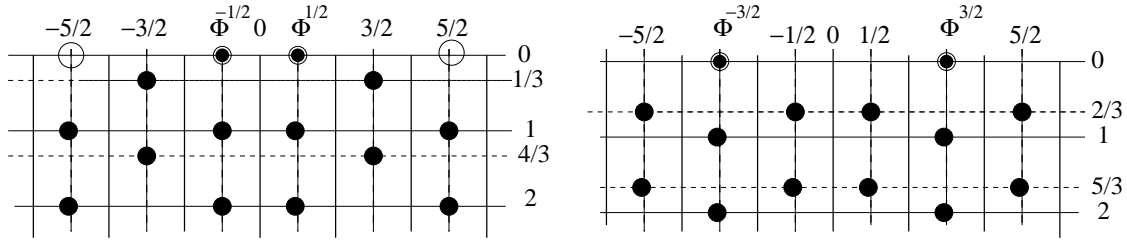}}
 \end{center}
 \protect\caption[3]{\label{fig3}$q=\pm 1/2$ and $q=\pm 3/2$ 
modules of the $Z_6$ theory.}
\end{figure}

As an example, we show in Fig.~\ref{fig3} the modules of
$\Phi^{\pm 1/2}$ and of $\Phi^{\pm 3/2}$ for the $Z_6$ theory.
Figs.~\ref{fig1} and \ref{fig3} taken together give the complete list of
modules for this theory.

In Appendix~\ref{secD211} we explicitly construct a
$(\Phi^{-1/2},\Phi^{+1/2})$ module (cf.~Fig.~\ref{fig3}) which is
degenerate on the first available descendent levels. The result for
the conformal dimension $\Delta_{\Phi^{1/2}}$ of the operators
$\Phi^{\pm 1/2}$ is in perfect agreement with the dimension of
the operator $\Phi_{(1)}\equiv\Phi_{(2,1,1)}$ of the elementary cell
(\ref{eq21}), the latter being computed with the appropriate boundary
term (\ref{eq28}). A simple analysis using the method of Weyl
reflections, based on Eq.~(\ref{eq35}), predicts that the module of
the operator $\Phi_{(2,1,1)}$ has to be degenerate twice at the level
1/3 and once at the level 1. This prediction is also confirmed by a
direct calculation in Appendix~\ref{secD211}.

The fact that the levels of degeneracy predicted by the reflection technique
agree with the outcome of direct calculations is important, even when the
result for the conformal dimension is trivial. Therefore, we expose briefly in
Appendix~\ref{secS111} the degeneracy calculation for the module 
of the identity operator $\Phi_{(0)} \equiv \Phi_{(1,1,1)} = I$.

Similarly, the direct calculation in Appendix~\ref{secD122} of degeneracies
for the module $(\Phi^{-1},\Phi^{1})$, cf.~Fig.~\ref{fig1}, provides an
agreement with the operator $\Phi_{(2)}\equiv\Phi_{(1,2,2)}$ of the elementary
cell (\ref{eq21}). (A technical point regarding this calculation is deferred
to Appendix~\ref{appmu3}.) It should be noticed that the calculations for the
operators $\Phi_{(1)}$ and $\Phi_{(2)}$ are actually generalised to the $Z_N$
theory with arbitrary even $N>6$. This is done towards the end of
Appendices~\ref{secD211}--\ref{secD122}.

Finally, the calculation in Appendix~\ref{sec_disorder} classifies the
operator $\Phi_{(1,2,1)}$ in Eq.~(\ref{eq22}) as the fundamental disorder
operator.

The direct degeneracy calculations of
Appendices~\ref{secS111}--\ref{sec_disorder} clearly support the
classification of sectors that we have proposed above. In particular,
Appendix~\ref{secD211} corroborates the argument in favour of
half-integer $q$.

On the other hand, it can be checked that once the
modules with half-integer $q$ have been admitted, the analysis using
the reflection method described after Eq.~(\ref{eq35}) makes the
assignment of the operators given above the only possible.

This is but with one correction.
Application of the reflection method alone cannot determine which
one of the operators $\Phi_{(1)}=\Phi_{(2,1,1)}$ and
$\Phi_{(2)}=\Phi_{(1,2,2)}$ should be related, respectively, to the doublets
$(\Phi^{-1/2},\Phi^{1/2})$ and $(\Phi^{-1},\Phi^{1})$. This is because the
boundary terms and the level spacings are the same in the above two sectors,
as witnessed by Eq.~(\ref{eq28}), and by Figs.~\ref{fig1} and \ref{fig3}.

To resolve this ambiguity we examine more closely the degeneracy structure of
the corresponding modules. According to the reflections, the module of the
operator $\Phi_{(2,1,1)}$ should be degenerate twice at the level 1/3 and once
at the level 1. Similarly, the module of the operator $\Phi_{(1,2,2)}$ should
be degenerate once at the level 1/3 and twice at the level 1. Using this
difference in the degeneracy patterns we could set up a simple test: we could
assume the assignment of $\Phi_{(2,1,1)}$ to the doublet
$(\Phi^{-1},\Phi^{1})$ and demand, accordingly, that the module of this
doublet be degenerate twice at the level 1/3. Calculation of the degeneracy
under this assumption, and for the level 1/3 alone, is very simple. It leads
immediately to a non-acceptable value of $\Delta$ for this doublet. We must
therefore conclude that $\Phi_{(2,1,1)}$ should be assigned to the doublet
$(\Phi^{-1/2},\Phi^{1/2})$ and the other operator, $\Phi_{(1,2,2)}$, to the
doublet $(\Phi^{-1},\Phi^{1})$.

Once the above ambiguity is resolved, with a slight intervention of the
degeneracy calculation, the assignment of the elementary cell operators and of
all their surrounding ghosts is uniquely defined.

We emphasise again that even though the boundary terms of the operators
$\Phi_{(1)}\equiv\Phi_{(2,1,1)}$ and $\Phi_{(2)}=\Phi_{(1,2,2)}$ happen to be
equal, their respective modules and the degeneracy patterns are different.
They therefore represent different sectors of the theory. In particular, this
difference is a strong argument against the would-be $Z_4$ symmetry of the
elementary cell that we tentatively discussed above.

So far we have not discussed the operator
$\Phi_{(3)}\equiv\Phi_{(1,3,1)}$, which is the last fundamental
operator in the list (\ref{eq21}), in the case of the $Z_6$ theory.
Its position in the elementary cell is shown in Fig.~\ref{fig2}. By
the analysis of reflections, cf.~Eq.~(\ref{eq35}), the module of this
operator should be degenerate twice at the level 2/3 and once at the
level 5/3. The direct calculation of the degeneracy at the level 5/3
is complicated. In general, the deeper the level, the more involved is
the calculation. This is because the number of states to be analysed
grows rapidly with the level. On the other hand, the assignment of
this operator to the sector $q=3/2$, the second module shown in
Fig.~\ref{fig3}, is the unique possibility allowed by the combined
argument of reflections [cf.~Eq.~(\ref{eq35})], the available modules
[cf.~Figs.~\ref{fig1} and \ref{fig3}], and the content of the
elementary cell [cf.~Fig.~\ref{fig2}]. There is however one important
difference with respect to the other operators having non-zero $q$
charge.  Namely, we shall argue below that despite the appearance of
Fig.~\ref{fig3}, $\Phi_{(1,3,1)}$ is {\em not} a doublet operator
$(\Phi^{-3/2},\Phi^{3/2})$, but rather another non-trivial singlet
operator, whose module is built on the state $\Phi^{3/2}$ alone at the
summit.

As has been shown above, the $Z_{N}$ charges of the primary fields
$\Phi^{q}$ are required to take half-integer values. This is because they have
been defined initially with respect to the $Z_{N}$ charges of the
parafermionic fields $\Psi^{k}$. For $\Psi^{k}$ we have admitted, initially,
the natural set of values $k=\pm 1,\pm 2,...,\pm(N/2-1),N/2$.

To obtain a more natural notation, and to bring out the structural
similarities of the final results for even and odd $N$, it is convenient to
redefine the initial $Z_{N}$ charges by multiplying them by two:
\beq
 Q=2q;\quad\Phi^{Q^{\star}}=\Phi^{q},
\label{eq39}
\eeq
in order that the new charges of the primary fields take the
natural set of values for the $Z_{N}$ group:
\beq
 Q=0,\pm 1,\pm 2,\ldots,\pm(N/2-1),N/2.
\label{eq40}
\eeq
Both definitions, $q$ and $Q$, have already been used in Ref.~\cite{ref2} for
the $Z_{N}$ theory with $N$ odd. For $N$ odd the difference between the two
notations is less pronounced and does not lead to the appearance of
half-integer values. As in Ref.~\cite{ref2}, we shall adopt a notation in
which the charges in the $Q$ notation are marked with an asterisk (we have
already tacitly done so in Eq.~(\ref{eq39})). For instance, when $N$ is even,
$\Phi^{1^{\star}}=\Phi^{\frac{1}{2}}$ will correspond to $\Phi$ with $Q=1$, or
$q=\frac{1}{2}$. When $N$ is odd \cite{ref2}, this same value, $Q=1$,
corresponds to the maximal possible $q$ charge, $q=\lfloor N/2 \rfloor$.

We can now complete the argument that $\Phi_{(1,3,1)}$ is in fact a
singlet.  As the $Z_{N}$ charges of the primary fields should be
defined modulo $N$ (once again, in order to be consistent with the
number of representations of the group $Z_N$), the charges $-N/2$ and
$N/2$ of a primary field $\Phi^q$ should be taken as equal.  This
implies that the corresponding operator is not a doublet
$(\Phi^{-N/2^\ast},\Phi^{N/2^\ast})$, but rather a singlet state
$\Phi^{N/2^\ast}$. In the special case of $Z_{6}$, this maximal $Q$
state will be the singlet $\Phi^{3^{\star}}\equiv
S^{3^{\star}}=\Phi^{3/2}$, shown in Fig.~\ref{fig2}.

Within the convention of doubled charges, we shall denote by $K$ the $Z_{N}$
charges of the parafermions:
\beq
 K=2k=\pm2,\pm4,...,\pm(N-2),N; \quad \Psi^{K^{\ast}}=\Psi^{k}.
\label{eq41}
\eeq
The $K$ charges then add up modulo $2N$. As a consequence, the $Z_N$
charges of the descendent fields will also be defined modulo $2N$. At
the same time, in every given module only $N$ distinct $Z_N$ charge
sectors will be occupied, as shown in Figs.~\ref{fig1} and \ref{fig3}.
This is consistent with the fact that the modules correspond to
representations of $Z_N$, whose number should be $N$, and not $2N$.

The above feature that the $Z_{N}$ charges of the parafermions take a
set of doubled values (as compared to charges of the primary
fields) is characteristic of a self-dual $Z_{N}$ theory. This is
explained in detail in Ref.~\cite{ref3}, within the context of the
parafermionic theory based on the {\em first solution} for the $Z_{N}$
chiral algebra. Note that in our construction self-duality has not
been assumed from the beginning, neither has it been used
anywhere. Rather, it has emerged by itself, in the process of
constructing a consistent theory.%
\footnote {In principle, one does not need the spin operator
$\sigma^{k}\equiv\Phi^{k}$ and its dual, the $Z_{N}$ disorder operator
$\mu^{k}\equiv\tilde{\Phi}^{k}$, to produce parafermions by taking products:
$\sigma^{k}\times\mu^{k}\sim\Psi^{2k}$. The parafermionic fields are produced
equally well in the products of different $Z_{2}$ disorder operators.
In Ref.~\cite{ref1} we have established the operator product
expansion $\Psi^{1}(z)R_{a}(0)\sim R_{a-2}(0)$, and used it to derive
characteristic equations. This relation implies, obviously, that $\Psi^{1}$ is
produced in the product $R_{a-2}(z)R_{a}(0)$.}

Before leaving the $q$ notation completely, we should remark that only in the
$Z_N$ theories with $N=2r$ and {\em even} $r$ is the notion of half-integer
charges $q$ mandatory. Namely, for any $r$, the doublet modules
$(\Phi^{-q},\Phi^q)$ are isomorphic to doublet modules
$(\Phi^{N/4-q},\Phi^{N/4+q})$ having their summits centered in $N/4$. For even
$r$, the charges of these two equivalent positionings of the module are either
both integer or both half-integer. For odd $r$, the charges of one of them is
integer and the other half-integer. In this latter case, one can therefore
avoid half-integer $q$ by taking one half of the doublet modules centered in
$0$ and one half centered in $N/4$.

The identification of the elementary cell operators (and of their ghosts),
which we have described above in some detail for the example of the $Z_{6}$
theory, are very similar in the $Z_{8}$ case. Like in the case of $Z_{6}$, the
application of the method of reflections leaves one ambiguity unresolved, that
of assigning the elementary cell operators $\Phi_{(2,1,1,1)}$ and
$\Phi_{(1,1,2,2)}$ to the sectors, respectively, of the doublets
$(\Phi^{-1/2},\Phi^{1/2})$ and $(\Phi^{-3/2},\Phi^{3/2})$, or in the opposite
order. This ambiguity is resolved in a way similar to the $Z_{6}$ case, by
demanding that the module of the doublet $(\Phi^{-3/2},\Phi^{3/2})$ be triply
degenerate on the first available level, which is $1/4$ in this case.
Similarly, this leads to a wrong value for $\Delta$.

For the $Z_{10}$ case additional ambiguities appear in the reflection method.
They could still all be resolved, this time with a more important intervention
of the direct degeneracy calculations. In particular, one would need to use
the results obtained in the Appendices \ref{secD211} and \ref{secD122} for the
modules $q=1/2$ and $q=1$, for general (even) $N$.

After these first cases of $Z_{6}$, $Z_{8}$ and $Z_{10}$ have been analysed
explicitly, the generalisation to higher values of $N$ is
straightforward. It is additionally checked by the results of Appendices
\ref{secD211} and \ref{secD122} for the doublets $q=1/2$ and 1, and by the
results of Appendix \ref{sec_disorder} for the disorder sector.

We could now summerise. The elementary cell of the $Z_{N}$ theory, with $N=2r$
even, is delimited by the inequality (\ref{eq20}). This is also the physical
domain of the $c=0$ theory. The symmetry of this cell, with respect to the
operator content of the $c=0$ theory, is $Z_{2}\times Z_{2}$. This is realised
by reflections in two orthogonal hyperplanes, defined as:
\beq
 \begin{tabular}{ll}
 \underline{First $Z_2$}: & $n'_{a}  \to n'_{a}$, \qquad $a=1,2,\ldots,r-2$ \\
                          & $n'_{r-1}\to n'_{r}$                            \\
                          & $n'_{r}  \to n'_{r-1}$                          \\
                          &                                                 \\
 \underline{Second $Z_2$}:& $n'_{1} \to 2r-n'_{1}-2\sum^{r-2}_{b=2}n'_{b}-
                             n'_{r-1}-n'_{r}$                               \\
                          & $n'_{a} \to n'_{a}$, \qquad $a=2,3,\ldots,r$    \\
 \end{tabular}
\label{eq42}
\eeq
The independent operators, after applying the above symmetry, are listed in
Eqs.~(\ref{eq21})--(\ref{eq22}). In this list there is precisely one operator
for each sector of the theory. Their identification is the following:
\bea
 \Phi_{(0)}  &\equiv& \Phi_{(1,1,1,\ldots,1,1,1)}=I=S^{0^{\ast}}\nn\\
 \Phi_{(1)}  &\equiv& \Phi_{(2,1,1,\ldots,1,1,1)}=D^{1^{\ast}}\nn\\
 \Phi_{(2)}  &\equiv& \Phi_{(1,2,1,\ldots,1,1,1)}=D^{2^{\ast}}\nn\\
 \ldots      &\ldots& \ldots \nn \\
 \Phi_{(r-2)}&\equiv& \Phi_{(1,1,1,\ldots,2,1,1)}=D^{r-2^{\ast}}\nn\\
 \Phi_{(r-1)}&\equiv& \Phi_{(1,1,1,\ldots,1,2,2)}=D^{r-1^{\ast}}\nn\\
 \Phi_{(r)}  &\equiv& \Phi_{(1,1,1,\ldots,1,3,1)}=S^{r^{\ast}}\nn\\
             &      & \Phi_{(1,1,1,\ldots,1,2,1)}=R
\label{eq43}
\eea
Here $S^{0^{\ast}}$ (charge $Q=0$) is a trivial singlet;
$D^{Q^{\ast}}$ (charge $Q=1,2,\ldots,r-1$) are $r-1$ different doublets;
note that $D^{Q^{\ast}}$ represents both components of the 
doublet $\left(\Phi^{-Q^{\ast}},\Phi^{Q^{\ast}}\right)$;
and $S^{r^{\ast}}$ (charge $Q=r$) is a nontrivial singlet. Finally,
$R$ is the $Z_{2}$ (charge conjugation) disorder operator.

The boundary terms for all these different sectors (i.e.,
the constant $B$ in the
Kac formula (\ref{eq8})) have been given in Eqs.~(\ref{eq28})--(\ref{eq29}).

When $c>0$, the unitary theories with integer $p>N-2$ have their physical
domains delimited by the inequalities (\ref{eq14})--(\ref{eq15}), which
replace the inequality (\ref{eq20}) of the $c=0$ theory.
The $Z_{2}\times Z_{2}$ symmetry of the physical domain is realised as:
\beq
 \begin{tabular}{ll}
 \underline{First $Z_2$}: & 
   \begin{tabular}{lll}
     $n'_a \to n'_a$,    & $n_a \to n_a$,    & $a=1,2,\ldots,r-2$ \\
     $n'_{r-1}\to n'_r$, & $n_{r-1} \to n_r$ &                    \\
     $n'_r \to n'_{r-1}$,& $n_r \to n_{r-1}$ &                    \\
   \end{tabular} \\
                         &                                        \\
 \underline{Second $Z_2$}:& $n'_{1} \to p+2-n'_{1}-2\sum^{r-2}_{b=2}n'_{b}-
                             n'_{r-1}-n'_{r}$                               \\
                          & $n_{1} \to p-n_{1}-2\sum^{r-2}_{b=2}n_{b}-
                             n_{r-1}-n_{r}$                                 \\
                          & $n'_{a} \to n'_{a}$, \qquad
                            $n_a \to n_a$, \qquad $a=2,3,\ldots,r$    \\
 \end{tabular}
\label{eq44}
\eeq
which replaces (\ref{eq42}).

The physical domain of a given $c>0$ unitary theory will contain more
(independent) operators than those listed in Eq.~(\ref{eq43}). Still, the
notion of the elementary cell remains significant. Firstly, because the list
(\ref{eq43}) presents all the fundamental operators for a given theory.
Secondly, the full set of operators delimited by the inequality (\ref{eq20})
(with trivial indices on the $\alpha_{+}$ side) still preserves the symmetry
(\ref{eq42}) {\em as far as the sector assignment is concerned}.
And thirdly, because the assignment of sector labels to all the operators in
the Kac table can be obtained by reflecting repeatedly the elementary
cell as we shall describe below.

\begin{figure}
\begin{center}
 \leavevmode
 \epsfysize=200pt{\epsffile{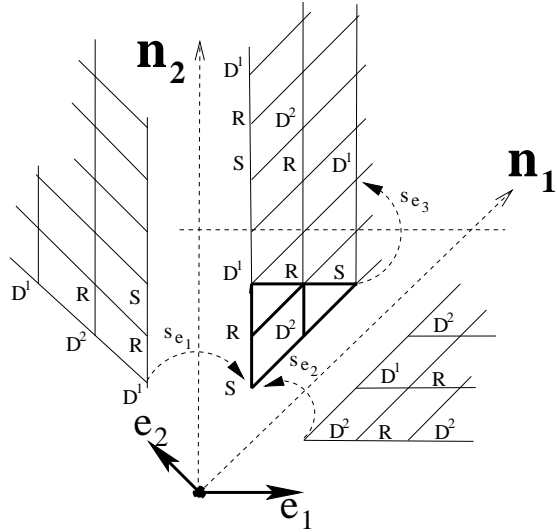}}
 \end{center}
 \protect\caption[3]{\label{fig4} The ghost environment of the
 elementary cell of the $Z_5$ theory is obtained by
 considering the reflections  $s_{\vec{e}_{1}}$, $s_{\vec{e}_{2}}$ and
 $s_{\vec{e}_{3}}$ (dashed arrows). Then, by using the fusions with the
 singlets $S$ in the figure, the whole basic layer can be filled.}
\end{figure}

The problem of filling the Kac table (i.e., assigning sector labels to
all the sites of the weight lattice) has already been dealt with in
Refs.~\cite{ref1,ref2}, for the theory with $N$ odd. The method was
based on Weyl reflections, evoking the assignment of sector labels to
the ghost operators, and on fusion of singlets $(q=0)$ with other
operators. This method can be simplified considerably by turning once
again to the trivial theory, $c=0$, which has the advantage that its
physical domain coincides with the elementary cell. Thus, all physical
operators lie in the ``basic layer'' \cite{ref2}, defined by the
absence of excitations on the $\alpha_{+}$ side.  The ghost
environment of the fundamental operators, the only physical ones of
the $c=0$ theory, is known from the preceding analysis. The rest of
the operators in the elementary cell are obtained by applying the
$Z_{2}\times Z_{2}$ symmetry to the fundamental ones,
cf.~Fig.~\ref{fig2} for the theory $Z_{6}$. Their ghost environment is
also obtained by this symmetry. To explain the resulting ghosts one
has to use all the simple reflections, including the reflection
(\ref{eq31a})--(\ref{eq31}) with respect to the direction
$\vec{e}_{r+1}$, for all the operators in the elementary cell.
The two dimensional basic layer
of the $Z_5$ theory \cite{ref1} 
is particularly adapted to show how the method 
of reflections  
 is applied. The elementary
 cell of this theory and the resulting 
 ghost environment is shown in Fig.~\ref{fig4}.
The resulting ghost environment of the elementary cell is sufficient to fill
the whole basic layer, by using fusions with the singlet ($q=0$) operators. As
in Ref.~\cite{ref2} we assume that the principal channel amplitudes are
non-vanishing in all fusions of singlets with other operators.

\begin{figure}
\begin{center}
 \leavevmode
 \epsfysize=200pt{\epsffile{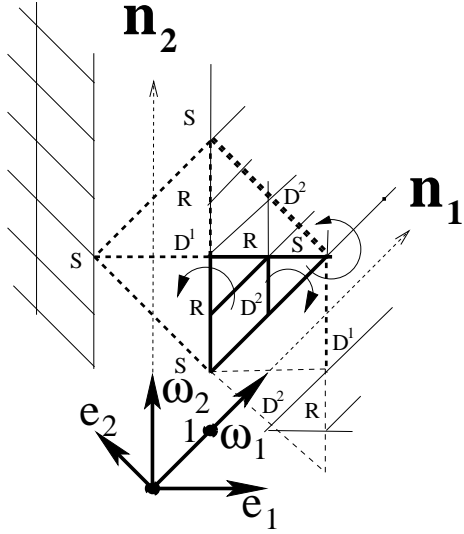}}
 \end{center}
 \protect\caption[3]{\label{fig5} The method of filling the lattice, by
 repeatedly reflecting the  
 elementary cell, is sketched for the case  of the $Z_5$ theory.}
\end{figure}

The final result of this analysis can be expressed in a much simpler way. Once
sector labels have been assigned to the operators of the elementary cell, the
assignment of the rest of the operators in the basic layer is obtained by
repeatedly reflecting the elementary cell in all its faces, filling
progressively in this way the whole lattice. In Fig.~\ref{fig5} we
show this procedure in the case of the 
$Z_5$ theory. 
The reflections employed in this process (henceforth referred to as Weyl
reflections of the second kind) are on a different footing than the
reflections evoking ghosts, which we have been discussing all the way before
(henceforth we refer to these as Weyl reflections of the first kind). In
particular, the reflections of the second kind are with respect to a given
face of the elementary cell, and not with respect to an intermediate plane as
in the case of reflections of the first kind,
cf.~Figs.~\ref{fig4}--\ref{fig5}. Also, the reflections of the second kind are
always applied to the elementary cell, even in the case of $c\neq 0$ theories,
while the reflections of the first kind are applied, in general, to the
physical domain of a given unitary theory. This can be seen from the
definitions (\ref{eq30})--(\ref{eq31}).

The second kind of reflections has no direct relation to the structure of
modules of primary operators. Its only significance is with respect to the
sector assignment.

We remark that the second kind of Weyl reflections appears also in a general
analysis of the coset-based conformal theories, with respect to the
distribution over the lattice of the boundary terms \cite{ref7,ref8}. The
methods we are using in this paper are quite different from those of
Refs.~\cite{ref7,ref8}. Also, the results that we have obtained are more
complete: we distinguish the different sectors not only by the actual values
of the boundary term (which in the present case of the $Z_N$ theory with $N$
even exhibits degeneracies), but also by the actual symmetry content (i.e., the
transformation properties under the group $Z_N$).

On the other hand, the demonstration of the coset-based results of
Refs.~\cite{ref7,ref8} is made using more general techniques,
and applies in a more general context. Consequently, the
convergence of the two approaches in the particular case of the $Z_N$ theory
can only be considered as being satisfactory and positive for both methods.

Saying it differently, the (second kind of) Weyl reflections symmetry
of the Kac table, established in Refs.~\cite{ref7,ref8}, in relation to
the values of the boundary terms, extends to the complete sector labeling
that we have found. One may hope that a complementary use of both methods
might prove to be useful in the future.

To complete our analysis of the sector labeling, it should be observed
that, as usual, having assigned the sector labels to the operators of
the basic layer, the labeling of the sites of the whole Kac table is
obtained by applying translations, as in
Refs.~\cite{ref1,ref2}. Evidently, the assignment of sector labels,
which we have been analysing above in the context of the $c=0$ theory,
holds for $c \neq 0$ theories in general.

The final result on the labeling of the Kac table can also be stated
algebraically, in a way similar to that in Ref.~\cite{ref2}. We use the
notation $\tilde{n}_{a}=|n_a-n'_a|$, and define the following quantities:
\bea
 x_a    &=& \tilde{n}_a, \qquad a=1,2,\ldots,r-2 \nn \\
 x_{r-1}&=& \tilde{n}_r, \nn \\
 x_r    &=& (\tilde{n}_{r-1}-\tilde{n}_r)/2.
\eea
The operator $\Phi_{(n_1,\ldots,n_r)(n'_1,\ldots,n'_r)}$ is then a disorder
operator if $x_r$ is non-integer. Otherwise, if $x_r$ is integer,
we have a singlet or a doublet operator with an associated charge $Q$.

This charge $Q$ can be written in several different ways. It can be noted
that $Q$ only depends on $\{x_a\}$ through $\tilde{x}_a=x_a \mbox{ mod } 2$.
We then have the recursive formula
\beq
 Q(\tilde{x}_1,\tilde{x}_2,\ldots,\tilde{x}_{k-1},1,0,\ldots,0) =
 k -  Q(\tilde{x}_1,\tilde{x}_2,\ldots,\tilde{x}_{k-1},0,0,\ldots,0), 
\eeq
with the initial condition $Q(0,\ldots,0) = 0$. This formula for $Q$
is actually not the simplest possible, but we have given it here because
it is identical to the one reported in Ref.~\cite{ref2} for the case
of $N$ odd. (Note however that in that case $\tilde{x}_{r-1}$ and
$\tilde{x}_r$ were defined in a different way.) A more appealing direct
formula is the following:
\beq
 Q(x_1,x_2,\ldots,x_r) =
 \sum_{a=1}^r \left[ \left( \sum_{b=a}^r x_b \right) \mbox{ mod } 2 \right].
 \label{filling}
\eeq

The sector labeling can alternatively be stated by giving the assignment
corresponding to the vector
\beq
 \sum_{a=1}^{r} \frac{\tilde{n}_{a}}{2} \vec{\omega}_{a}
 \label{omega_vec}
\eeq
with respect to an orthonormal (hypercubic) basis, rather than with respect
to the Dynkin labels, cf.~Fig.~\ref{fig2} for the $r=3$ case.
This basis can be chosen such that
\bea
 \vec{\omega}_a &=& (1,\ldots,1,0,\ldots,0), \qquad a=1,2,\ldots,r-2 \nn \\
 \vec{\omega}_{r-1} &=& (1/2,\ldots,1/2,-1/2), \nn \\
 \vec{\omega}_r &=& (1/2,\ldots,1/2).
\eea
In the first line, the first $a$ entries of $\vec{\omega}_a$ are $1$.
Note that this choice satisfies the scalar products (\ref{eq23}).
Now let $y_a$ be the coordinates of the vector (\ref{omega_vec})
with respect to this basis. The periodicity of the assignment is such
that the sector label only depends on the values of
$\tilde{y}_a = y_a \mbox{ mod } 1$. The result is the following:
\begin{itemize}
 \item If $Q$ of the coordinates $\{ \tilde{y}_a \}$ are equal to $1/2$
       and the remaining $r-Q$ coordinates are zero, we have a
       singlet/doublet operator of charge $Q$. In other words,
       $Q=2\sum_{a=1}^r \tilde{y}_a$ in this case.
       Note in particular that there are ${r \choose Q}$ charge $Q$
       operators within each cubic unit cell.
 \item If each of the $\{ \tilde{y}_a \}$ is either $1/4$ or $3/4$,
       we have a disorder operator $R$. In particular, there are
       $2^r$ disorder operators within each cubic unit cell.
\end{itemize}
Let us finally remark that this same result holds true in the $B_r$ algebra
case (with $N=2r+1$ odd), provided that one chooses the basis
\bea
 \vec{\omega}_a &=& (1,\ldots,1,0,\ldots,0), \qquad a=1,2,\ldots,r-1 \nn \\
 \vec{\omega}_r &=& (1/2,\ldots,1/2).
\eea

\appendix

\def\theequation{\thesection.\arabic{equation}}

\section{Singlet $q=0$}
\setcounter{equation}{0}
\label{secS111}

We here expose the degeneracy calculation of the fundamental singlet $S^0$ in
the $Z_6$ theory. According to the technique of Weyl reflections, the
corresponding module should be three times degenerate at level $2/3$.

Extensive use of commutation relations shows that the unreduced module
has three states at level $2/3$:
\beq
 A^1_{-\frac23} \Phi^0, \qquad
 A^2_0 A^{-1}_{-\frac23} \Phi^0, \qquad
 A^2_{-\frac23} \Phi^0.
 \label{3states}
\eeq
Note that the second of these states exploits the zero mode $A^2_0$
at level $2/3$. We have checked that repeated actions of the various
zero modes at level $2/3$ do not produce additional independent states.

In order to have the required three degeneracies, we impose the complete
degeneracy of the states (\ref{3states}). This results in the following
three independent conditions:
\bea
 A^{-1}_{\frac23} \left( A^1_{-\frac23} \Phi^0 \right) &=& 0,
 \label{3cond1} \\
 A^{-1}_{\frac23} \left( A^2_0 A^{-1}_{-\frac23} \Phi^0 \right) &=& 0,
 \label{3cond2} \\
 A^{-1}_{\frac23} A^{-1}_0 \left( A^2_{-\frac23} \Phi^0 \right) &=& 0.
 \label{3cond3}
\eea
The first condition, Eq.~(\ref{3cond1}), can be rewritten through
\beq
 A^{-1}_{\frac23} A^1_{-\frac23} \Phi^0 = \frac{10 \Delta_{\Phi^0}}{3c} \Phi^0,
\eeq
fixing $\Delta_{\Phi^0} = 0$; this was expected since the most relevant singlet
must be the identity operator $\Phi_{(1,1,1)}$.
The second condition, Eq.~(\ref{3cond2}),
further implies the vanishing of some matrix elements at level $1$:
\beq
 A^{-1}_{\frac23} A^1_{\frac13} A^1_{-\frac13} A^{-1}_{-\frac23} =
 A^1_{\frac23} A^{-1}_{\frac13} A^1_{-\frac13} A^{-1}_{-\frac23} = 0.
\eeq
Finally, the third condition, Eq.~(\ref{3cond3}), can be written as
\beq
 A^{-2}_{\frac23} A^2_{-\frac23} \Phi^0 = 0.
\eeq
Exploiting the commutation relation
\beq
 \left( A^{-2}_{\frac53} A^2_{-\frac53} - \frac73 A^{-2}_{\frac23}
 A^2_{-\frac23} + A^2_{\frac23} A^{-2}_{-\frac23} \right) \Phi^0 =
 \frac{16 \Delta_{\Phi^0}}{3c} \Phi^0,
\eeq
this means that we have $A^{-2}_{\frac53} A^2_{-\frac53} \Phi^0 = 0$
as well.

\section{Doublet $q=\pm 1/2$}
\setcounter{equation}{0}
\label{secD211}

According to the reflection method, the doublet
$D^{\pm 1/2} = \Phi_{(2,1,1)}$ of the $Z_6$ theory is doubly degenerate
at level $1/3$ and degenerate once at level $1$.

We begin by investigating the $q=3/2$ sector at level $1/3$.
There is just a single state:
\beq
 A^1_{-\frac13} \Phi^{1/2} \propto A^2_{-\frac13} \Phi^{-1/2},
\eeq
and we require it to be completely degenerate. This amounts to imposing
the condition
$A^{-1}_{\frac13} A^{1}_{-\frac13} \Phi^{1/2} = 0$,
which, by the use of commutation relations, can be rewritten as
\beq
 \left( -\frac19 + \frac{10}{3c} \Delta_{\Phi^{1/2}}
  - A^{1}_0 A^{-1}_0 \right) \Phi^{1/2} = 0.
\eeq
This fixes the zero-mode eigenvalue, defined by
$A^1_0 \Phi^{-1/2} \equiv h \Phi^{1/2}$, as
\beq
 h^2 = -\frac19 + \frac{10}{3c} \Delta_{\Phi^{1/2}}.
 \label{fixtheta}
\eeq

Imposing also the conjugate degeneracy in the $q=-3/2$ sector at level
$\frac13$, we can now reduce the module by setting
$A^{\pm 1}_{-\frac13} \Phi^{\pm 1/2} = 0$.
The reduced module is then completely empty at level $1/3$.

The next available level is level $1$, and we here examine the sector
$q=1/2$ (the case $q=-1/2$ is equivalent, by charge conjugation symmetry).
In the reduced module, there are only two independent ways of
descending to this position, and we consider a linear combination of
the corresponding two states:
\beq
 \chi^{1/2}_{-1} = a L_{-1} \Phi^{1/2} + b A^{1}_{-1} \Phi^{-1/2}.
 \label{q1/2state}
\eeq
We impose degeneracy of $\chi^{1/2}_{-1}$ by demanding that
$L^1 \chi^{1/2}_{-1} = A^{-1}_1 \chi^{1/2}_{-1} = 0$.
In terms of the matrix elements defined by
\bea
 \mu_{11} \Phi^{1/2}  &\equiv& L_1 L_{-1} \Phi^{1/2},
 \label{matel1a} \\
 \mu_{12} \Phi^{1/2}  &\equiv& L_1 A^{1}_{-1} \Phi^{-1/2}, \\
 \mu_{21} \Phi^{-1/2} &\equiv& A^{-1}_1 L_{-1} \Phi^{1/2}, \\
 \mu_{22} \Phi^{-1/2} &\equiv& A^{-1}_1 A^{1}_{-1} \Phi^{-1/2},
 \label{matel1d}
\eea
the degeneracy constraint can be written
\beq
 \mu_{11} \mu_{22} - \mu_{12} \mu_{21} = 0.
 \label{degD2q1}
\eeq

The four matrix elements are easily evaluated:
\bea
 \mu_{11} &=& 2 \Delta_{\Phi^{1/2}},
 \label{matel1asol} \\
 \mu_{12} = \mu_{21} &=& \Delta_1 h,
 \label{matel1bcsol} \\
 \mu_{22} &=& \frac13 h^2 + \frac29 + \frac{10}{3c} \Delta_{\Phi^{1/2}}.
 \label{matel1dsol}
\eea
Inserting this in Eq.~(\ref{degD2q1})---and using also Eq.~(\ref{fixtheta})
for $h^2$---we obtain the physical solutions
\beq
 \Delta_{\Phi^{1/2}} = \frac{1}{48} \left( 25 - c \pm
 \sqrt{(c-5)(c-125)} \right).
 \label{solD211}
\eeq

The calculation can readily be generalised from the $Z_6$ theory to
the case of $Z_N$ with $N \ge 6$ even. The complete degeneracy in the
$q = \pm 3/2$ sector takes the form
$A^{\pm 1}_{-1+4/N} \Phi^{\pm 1/2} = 0$ and leads to
\beq
 h^2 = - \frac{N-2}{N^2} + \frac{4(N-1)}{N c} \Delta_{\Phi^{1/2}}.
\eeq

Following Ref.~\cite{ref2} we now assume that $N/2-3$ further degeneracies
can be imposed at the levels predicted by the Weyl reflections,
without fixing $\Delta_{\Phi^{1/2}}$, and in such a way that
the reduced module will be completely empty at all levels strictly between
0 and 1. The correctness of this assumption can be explicitly checked for
the first few values of $N$.

Turning to level 1, we once again impose the degeneracy of the state
(\ref{q1/2state}). The matrix elements (\ref{matel1asol})--(\ref{matel1bcsol})
are unchanged, whereas
\beq
 \mu_{22} = \frac{N-4}{N} h^2 + \frac{N+2}{N^2} + \frac{4(N-1)}{N c}
 \Delta_{\Phi^{1/2}}.
\eeq
Solving Eq.~(\ref{degD2q1}) we obtain the solutions
\beq
 \Delta_{\Phi^{1/2}} = \frac{1}{2N(N-2)} \left( (N-1)^2 - c \pm
 \sqrt{\left( c-(N-1) \right) \left(c - (N-1)^3 \right)} \right).
\eeq
generalising Eq.~(\ref{solD211}). This corresponds to the dimension
of the operator $\Phi_{(2,1,1,\ldots,1,1)}$.

\section{Doublet $q=\pm 1$}
\setcounter{equation}{0}
\label{secD122}

In the $Z_6$ case, which we consider first, the method of Weyl reflections
indicates the existence of a doublet $\Phi_{(1,2,2)}$ with one degeneracy
at level $1/3$ and two degeneracies at level $1$. We here show that this is
the doublet $D^1$.

First, to obtain a consistent doublet module, we need to make sure that
$A^1_0 \Phi^1 = 0$. In particular, this implies $A^{-1}_0 A^1_0 \Phi^1 = 0$,
and using the commutation relation $\{ \Psi^1,\Psi^{-1} \} \Phi^1$ we obtain
\beq
 A^1_{\frac13} A^{-1}_{-\frac13} \Phi^1 =
 \left( -\frac19 + \frac{10}{3c} \Delta_{\Phi^1} \right) \Phi^1.
 \label{mu1-1}
\eeq
Note that this consistency requirement does not count as a degeneracy within
the reflection method.

At level $1/3$, the $q=3$ sector is now empty. This follows from
$\{ \Psi^1,\Psi^1 \} \Phi^1$:
\beq
 A^1_{-\frac13} A^1_0 \Phi^1 = \frac12 \lambda^{1,1}_2 A^2_{-\frac13} \Phi^1
 = 0.
\eeq
On the other hand, the two states at level $1/3$ in the $q=0$ sector
can be used to form a singular state:
\beq
 \chi^0_{-\frac13} = a A^1_{-\frac13} \Phi^{-1} + b A^{-1}_{-\frac13} \Phi^1,
 \label{constab}
\eeq
for some constants $a$ and $b$.
Imposing the constraints $A^{\pm 1}_{\frac13} \chi^0_{-\frac13}$,
and defining the matrix elements
\bea
 \mu_{1,1}  \Phi^1 &\equiv& A^1_{\frac13} A^1_{-\frac13} \Phi^{-1}, \\
 \mu_{1,-1} \Phi^1 &\equiv& A^1_{\frac13} A^{-1}_{-\frac13} \Phi^1,
\eea
the degeneracy condition takes the form
\beq
 \mu_{1,1} = \pm \mu_{1,-1}.
\eeq

The matrix element $\mu_{1,-1}$ is fixed by Eq.~(\ref{mu1-1}), and
$\mu_{1,1}$ follows from $\{ \Psi^1,\Psi^1 \} \Phi^{-1}$:
\beq
 A^1_{\frac13} A^1_{-\frac13} \Phi^{-1} = \lambda^{1,1}_2 A^2_0 \Phi^{-1}
 \equiv \lambda^{1,1}_2 h \Phi^1,
\eeq
where $h$ is the zero-mode eigenvalue.
The degeneracy criterion then reads
\beq
 \lambda^{1,1}_2 h = \pm \left(-\frac19 + \frac{10}{3c} \Delta_{\Phi^1}\right).
 \label{deglev13}
\eeq
In the corresponding reduced module there is now only one state at
level $1/3$:
\beq
 A^1_{-\frac13} \Phi^{-1} = A^{-1}_{-\frac13} \Phi^1.
\eeq

At level $1$, we can now choose to impose a further degeneracy either
in the $q=1$ sector or in the $q=2$ sector (the cases $q=-1$ and $q=-2$
are equivalent by $Z_N$ charge conjugation). Extensive use of the commutation
relations reveals that among the many possible ways of descending to level 1
(including the use of zero modes at level 1), only a few are independent.
More precisely, in the $q=1$ sector only the two states
\beq
 L_{-1} \Phi^1 \mbox{ and } A^1_{-\frac23} A^{-1}_{-\frac13} \Phi^1
 \label{statesq1}
\eeq
are linearly independent. Likewise, in the $q=2$ sector we have the two
independent states
\beq
 A^1_{-1} \Phi^1 \mbox{ and } A^2_{-\frac23} A^{-1}_{-\frac13} \Phi^1.
 \label{statesq2}
\eeq

We first consider the possibility of degenerating a linear combination
of the states (\ref{statesq1}) in the $q=1$ sector. Setting
\beq
 \chi^1_{-1} = a L_{-1} \Phi^1 + b A^1_{-\frac23} A^{-1}_{-\frac13} \Phi^1,
 \label{singstatchi1-1}
\eeq
we require $L_1 \chi^1_{-1} = 0$ and $A^{-1}_{\frac23} \chi^1_{-1} = 0$.
(The constants $a$ and $b$ are not related to those in Eq.~(\ref{constab}).)
In terms of the matrix elements
\bea
 \mu_0 \Phi^1 &\equiv& L_1 L_{-1} \Phi^1, \\
 \mu_1 \Phi^1 &\equiv& L_1 A^1_{-\frac23} A^{-1}_{-\frac13} \Phi^1,
 \label{defmu1} \\
 \mu_2 \left( A^{-1}_{-\frac13} \Phi^1 \right) &\equiv&
 A^{-1}_{\frac23} L_{-1} \Phi^1,
 \label{defmu2} \\
 \mu_3 \left( A^{-1}_{-\frac13} \Phi^1 \right) &\equiv&
 A^{-1}_{\frac23} A^1_{-\frac23} A^{-1}_{-\frac13} \Phi^1,
 \label{defmu3}
\eea
the degeneracy criterion reads
\beq
 \mu_0 \mu_3 - \mu_1 \mu_2 = 0.
 \label{deglev1q1}
\eeq

The evaluation of the first three matrix elements is straightforward:
\bea
 \mu_0 &=& 2 \Delta_{\Phi^1}, \\
 \mu_1 &=& \frac43 \left( -\frac19 + \frac{10}{3c} \Delta_{\Phi^1} \right), \\
 \mu_2 &=& \frac43.
\eea
The last one follows from the commutation relation
\beq
 \left( A^{-1}_{\frac23} A^1_{-\frac23} -
 \frac13 A^{-1}_{-\frac13} A^1_{\frac13} + A^1_{-\frac13} A^{-1}_{\frac13}
 \right) A^{-1}_{-\frac13} \Phi^1 = \frac{10}{3c} \left( \Delta_{\Phi^1} +
 \frac13 \right) A^{-1}_{-\frac13} \Phi^1,
\eeq
which gives
\beq
 \mu_3 = \frac13 \left( -\frac19 + \frac{10}{3c} \Delta_{\Phi^1} \right) -
 \lambda^{1,1}_{2} h + \frac{10}{3c} \left( \Delta_{\Phi^1} + \frac13 \right).
\eeq

Inserting this in Eq.~(\ref{deglev1q1}) one obtains a quadratic equation for
$\Delta_{\Phi^1}$, whose solutions depend on the sign chosen in
Eq.~(\ref{deglev13}). With the plus sign, we obtain the physically
acceptable solutions
\beq
 \Delta_{\Phi^1} = \frac{1}{30} \left( 25 - c \pm \sqrt{(c-5)(c-125)} \right),
 \label{solD122}
\eeq
whilst the minus sign leads to the unacceptable solutions
\beq
 \Delta_{\Phi^1} = \frac{1}{210} \left( 25+2c\pm\sqrt{4c^2-460c+625} \right).
\eeq

Instead of imposing a degeneracy at level 1 in the $q=1$ sector, one could
have chosen to degenerate a linear combination of the states (\ref{statesq2})
in the $q=2$ sector. Note that the symmetry of the unreduced module is such
that the $q=1$ and $q=2$ sectors are equivalent. However, in the construction
of the doublet module we have broken that symmetry by placing the summits in
the $q=1$ sector. There is thus a priori no reason to expect that these two
sectors should be equivalent as far as the submodules are concerned. And
indeed we have seen that at level $1/3$ there are zero states in the $q=3$
sector and two states in the $q=0$ sector (of which one has been degenerated).

We therefore consider the singular vector
\beq
 \chi^2_{-1} = a A^1_{-1} \Phi^1 + b A^2_{-\frac23} A^{-1}_{-\frac13} \Phi^1,
\eeq
subject to the constraints $A^{-1}_1 \chi^2_{-1} = 0$ and
$A^{-2}_{\frac23} \chi^2_{-1} = 0$. Defining the matrix elements
\bea
 \tilde{\mu}_0 \Phi^1 &\equiv& A^{-1}_1 A^1_{-1} \Phi^1, \\
 \tilde{\mu}_1 \Phi^1 &\equiv& A^{-1}_1 A^2_{-\frac23} A^{-1}_{-\frac13}
 \Phi^1, \\
 \tilde{\mu}_2 \left( A^{-1}_{-\frac13} \Phi^1 \right) &\equiv&
 A^{-2}_{\frac23} A^1_{-1} \Phi^1, \\
 \tilde{\mu}_3 \left( A^{-1}_{-\frac13} \Phi^1 \right) &\equiv&
 A^{-2}_{\frac23} A^2_{-\frac23} A^{-1}_{-\frac13} \Phi^1,
 \label{tildemu3}
\eea
the degeneracy criterion is
\beq
 \tilde{\mu}_0 \tilde{\mu}_3 - \tilde{\mu}_1 \tilde{\mu}_2 = 0.
 \label{deglev1q2}
\eeq

Once again, the evaluation of the first three matrix elements is
straightforward:
\bea
 \tilde{\mu}_0 &=& \frac29 + \frac{10}{3c} \Delta_{\Phi^1}, \\
 \tilde{\mu}_1 &=& \lambda^{1,1}_2 h^2 + \frac35 \lambda^{-1,2}_1
 \left( -\frac19 + \frac{10}{3c} \Delta_{\Phi^1} \right), \\
 \tilde{\mu}_2 &=& h + \frac35 \lambda^{-1,2}_1.
\eea
Evaluating $\tilde{\mu}_3$ is considerably more involved, calling for the use
of numerous commutation relations (see Appendix~\ref{appmu3}). The result is
\beq
 \tilde{\mu}_3 = h^2 + \frac{\lambda^{-1,2}_1}{\lambda^{1,1}_2} \left(
 -\frac{1}{45} + \frac{8 \Delta_{\Phi^1}}{3c} \right) +
 \frac{\lambda^{1,2}_3 \lambda^{1,3}_{-2}}{2 \lambda^{1,1}_2}
 \left( h + \frac{\lambda^{-1,2}_1}{5} \right) -
 \frac{\lambda^{-1,2}_1}{5} \left( h + \frac{3 \lambda^{-1,2}_1}{5} \right).
 \label{result_tildemu3}
\eeq

Inserting these matrix elements---and the values of the various
structure constants---in Eq.~(\ref{deglev1q2}), one finds that the
physically acceptable solutions for $\Delta_{\Phi^1}$ are once again
given by Eq.~(\ref{solD122}).

According to the reflection technique, there should be a second degeneracy
at level 1, apart from the one that we have imposed either in the $q=1$ or
in the $q=2$ sector (with identical results). This second degeneracy will
not have any consequence for the determination of $\Delta_{\Phi^1}$ or of
$h$, but will ensure that the ghost state at level $1$ has a valid
doublet structure. In other words, it imposes a constraint on the module
of the ghost operator at level 0; this is analoguous to the constraint
$A^1_0 \Phi^1 = 0$, discussed above Eq.~(\ref{mu1-1}), which we have imposed
on the module of the physical operator $\Phi_{(1,2,2)}$.

The result (\ref{solD122}) can be generalised to the case of $Z_N$ with
$N>6$ even, as follows. We begin by imposing the full degeneracy of the
first descendent $q=2$ state, which is $A^1_{-1+6/N} \Phi^1$. The
constraint $A^{-1}_{1-6/N} A^1_{-1+6/N} \Phi^1$ then leads to the fixation
of the matrix element
\beq
 A^1_{2/N} A^{-1}_{-2/N} \Phi^1 = - \frac{2(N-4)}{N^2} + \frac{4(N-1)}{N c}
 \Delta_{\Phi^1}.
\eeq
This generalises Eq.~(\ref{mu1-1}), but note that here it counts as a
genuine degeneracy, rather than just a consistency requirement, as was
the case for $N=6$.

Imposing next a degeneracy in the $q=0$ sector at level $2/N$ will fix
the zero mode eigenvalue:
\beq
 \lambda_2^{1,1} h = \pm \left( - \frac{2(N-4)}{N^2} + \frac{4(N-1)}{N c}
 \Delta_{\Phi^1} \right).
 \label{deglev13gen}
\eeq
This is completely analogous to Eq.~(\ref{deglev13}).

Taking our cue from Appendix~\ref{secD211} (and from Ref.~\cite{ref2}) we now
assume that $N/2-4$ further degeneracies can be imposed at the levels
predicted by the Weyl reflections, without fixing $\Delta_{\Phi^1}$, and in
such a way that the reduced module will be completely empty at all levels
strictly between 0 and 1 in the sectors $q \neq 0$. However, note that the
criterion (\ref{deglev13gen}) still leaves one state at level $2/N$ in the
$q=0$ sector.

Finally we demand the degeneracy of
\beq
 \chi^1_{-1} = a L_{-1} \Phi^1 + b A^1_{-1+2/N} A^{-1}_{-2/N} \Phi^1
\eeq
cf.~Eq.~(\ref{singstatchi1-1}). Defining $\mu_1$, $\mu_2$ and $\mu_3$
as the obvious generalisations of Eqs.~(\ref{defmu1})--(\ref{defmu3}),
the commutation relations lead to
\bea
 \mu_1 &=& \frac{2(N-2)}{N} \left( - \frac{2(N-4)}{N^2} +
   \frac{4(N-1)}{N c} \Delta_{\Phi^1} \right), \\
 \mu_2 &=& \frac{2(N-2)}{N}, \\
 \mu_3 &=& \frac{N-4}{N} \left( - \frac{2(N-4)}{N^2} +
   \frac{4(N-1)}{N c} \Delta_{\Phi^1} \right) - \lambda_2^{1,1} h +
   \frac{4(N-1)}{N c} \left( \Delta_{\Phi^1} + \frac{2}{N} \right).
\eea
The condition (\ref{deglev1q1}) then yields the physically acceptable
solution
\beq
 \Delta_{\Phi^1} = \frac{1}{N(N-1)} \left( (N-1)^2 - c \pm
   \sqrt{\left( c-(N-1) \right) \left(c-(N-1)^3 \right)} \right),
\eeq
generalising Eq.~(\ref{solD122}). This corresponds to the dimension
of the operator $\Phi_{(1,2,1,\ldots,1,1)}$ for $N>6$, and to
$\Phi_{(1,2,2)}$ for $N=6$.

\section{Calculation of $\tilde{\mu}_3$}
\setcounter{equation}{0}
\label{appmu3}

We wish to evaluate the matrix element $\tilde{\mu}_3$ defined
by Eq.~(\ref{tildemu3}). In the following we shall make extensive use of
various commutation relations; which relations are being used can be inferred
from the presence of the corresponding structure constants. We first have:
\bea
 \tilde{\mu}_3 \left( A^{-1}_{-\frac13} \Phi^1 \right) &\equiv&
 A^{-2}_{\frac23} A^2_{-\frac23} A^{-1}_{-\frac13} \Phi^1 \nn \\
 &=& A^{-2}_{\frac23} \left( A^{-1}_{\frac13} A^2_{-\frac43} -
 \frac15 \lambda^{-1,2}_1 A^1_{-1} \right) \Phi^1.
 \label{3star}
\eea
The first term of the right-hand side of Eq.~(\ref{3star}) also enters
in:
\beq
 \left( A^{-2}_{\frac23} A^{-1}_{\frac13}
 - A^{-1}_{-\frac13} A^{-2}_{\frac43} \right)
 A^2_{-\frac43} \Phi^1 = \frac13 \lambda^{1,2}_3 A^3_1 A^2_{-\frac43} \Phi^1,
 \label{2star}
\eeq
where we have used that $A^{-1}_{\frac43} A^2_{-\frac43} \Phi^1 = 0$.
To simplify Eq.~(\ref{2star}) we shall use that
\beq
 \lambda^{1,1}_2 A^2_{-\frac43} \Phi^1 = A^1_{-\frac13} A^1_{-1} \Phi^1.
 \label{factorize}
\eeq
The second term on the left-hand side of Eq.~(\ref{2star}) contains
the piece $A^{-2}_{\frac43} A^2_{-\frac43} \Phi^1$, which also enters in:
\beq
 \left( A^{-2}_{\frac43} A^1_{-\frac13}
 - A^1_{\frac13} A^{-2}_{\frac23} \right) A^1_{-1} \Phi^1 =
 \frac15 \lambda^{-1,2}_1 A^{-1}_1 A^1_{-1} \Phi^1,
 \label{1star}
\eeq
where we have used that $A^1_{\frac23} A^1_{-1} \Phi^1 = 0$.

Eq.~(\ref{1star}) calls for the evaluation of
$A^{-2}_{\frac23} A^1_{-1} \Phi^1$, which is also required to get the second
term on the right-hand side of Eq.~(\ref{3star}). We have
\beq
 \left( A^{-2}_{\frac23} A^1_{-1} - A^1_{-\frac13} A^{-2}_0 \right) \Phi^1
 = \frac35 \lambda^{-1,2}_1 A^{-1}_{-\frac13} \Phi^1,
\eeq
where we have used that $A^1_0 \Phi^1 = 0$. Thus
\beq
 A^{-2}_{\frac23} A^1_{-1} \Phi^1 = h A^1_{-\frac13} \Phi^{-1} +
 \frac35 \lambda^{-1,2}_1 A^{-1}_{-\frac13} \Phi^1.
 \label{usein3star}
\eeq
Acting on this with $A^1_{\frac13}$ we find
\beq
 A^1_{\frac13} A^{-2}_{\frac23} A^1_{-1} \Phi^1 =
 h^2 \lambda^{1,1}_2 \Phi^1 + \frac35 \lambda^{-1,2}_1
 \left( -\frac19 + \frac{10}{3c} \Delta_{\Phi^1} \right) \Phi^1,
 \label{bpiece}
\eeq
which is the second term on the left-hand side of Eq.~(\ref{1star}).

To obtain the right-hand side of Eq.~(\ref{1star}) we use that
\beq
 A^{-1}_1 A^1_{-1} \Phi^1 = \left( \frac29 + \frac{10}{3c} \Delta_{\Phi^1}
 \right) \Phi^1.
 \label{cpiece}
\eeq
Rearranging Eq.~(\ref{1star}), and inserting
Eqs.~(\ref{bpiece})--(\ref{cpiece}), one finds that
\beq
 A^{-2}_{\frac43} A^1_{-\frac13} A^1_{-1} \Phi^1 = \left(
 h^2 \lambda^{1,1}_2 + \frac35 \left( -\frac19 + \frac{10}{3c} \Delta_{\Phi^1}
 \right) \lambda^{-1,2}_1 + \frac15 \left( \frac29 + \frac{10}{3c}
 \Delta_{\Phi^1} \right) \lambda^{-1,2}_1 \right) \Phi^1.
\eeq
Using Eq.~(\ref{factorize}), and doing some simplification, we arrive
at
\beq
 A^{-2}_{\frac43} A^2_{-\frac43} \Phi^1 = \left(
 h^2 + \frac{\lambda^{-1,2}_1}{\lambda^{1,1}_2} \left(
 -\frac{1}{45} + \frac{8}{3c} \Delta_{\Phi^1} \right) \right) \Phi^1,
 \label{usein2star}
\eeq
which is essentially the second term on the left-hand side of
Eq.~(\ref{2star}).

It remains to get the right-hand side of Eq.~(\ref{2star}).
Using again (\ref{factorize}), we have
\beq
 A^3_1 A^2_{-\frac43} \Phi^1 = \frac{1}{\lambda^{1,1}_2}
 A^3_1 A^1_{-\frac13} A^1_{-1} \Phi^1,
 \label{4star1}
\eeq
which can be simplified through
\beq
 \left( A^3_1 A^1_{-\frac13} - A^1_{-\frac13} A^3_1 \right) A^1_{-1} \Phi^1
 = \frac12 \lambda^{1,3}_{-2} A^{-2}_{\frac23} A^1_{-1} \Phi^1
 \label{4star2}
\eeq
To make this explicit, we need $A^3_1 A^1_{-1} \Phi^1$
and $A^{-2}_{\frac23} A^1_{-1} \Phi^1$. The first of these reads
\beq
 A^3_1 A^1_{-1} \Phi^1 = \lambda^{1,3}_{-2} A^{-2}_0 \Phi^1
 = h \lambda^{1,3}_{-2} \Phi^{-1},
\eeq
whilst the second is obtained from
\beq
 \left( A^{-2}_{\frac23} A^1_{-1} - A^1_{-\frac13} A^{-2}_0 \right) \Phi^1
 = \frac35 \lambda^{-1,2}_1 A^{-1}_{-\frac13} \Phi^1,
\eeq
yielding
\beq
 A^{-2}_{\frac23} A^1_{-1} \Phi^1 = h A^1_{-\frac13} \Phi^{-1} +
 \frac35 \lambda^{-1,2}_1 A^{-1}_{-\frac13} \Phi^1.
\eeq
Inserting these bits and pieces in Eqs.~(\ref{4star1})--(\ref{4star2})
we get
\beq
 A^3_1 A^2_{-\frac43} \Phi^1 = \frac{1}{\lambda^{1,1}_2} \left(
 \frac32 h \lambda^{1,3}_{-2} A^1_{-\frac13} \Phi^{-1} + \frac{3}{10}
 \lambda^{1,3}_{-2} \lambda^{-1,2}_1 A^{-1}_{-\frac13} \Phi^1 \right).
 \label{usein2starbis}
\eeq

Finally, we inject Eqs.~(\ref{usein2star}) and (\ref{usein2starbis})
in Eq.~(\ref{2star}) in order to isolate
$A^{-2}_{\frac23} A^{-1}_{\frac13} A^2_{-\frac43} \Phi^1$.
Inserting this, and also Eq.~(\ref{usein3star}), in Eq.~(\ref{3star})
we infer the value of $\tilde{\mu}_3$ given in
Eq.~(\ref{result_tildemu3}).

\section{Disorder operators}
\setcounter{equation}{0}
\label{sec_disorder}

The general properties of the disorder operators (fusion rules,
analytic continuation, etc.) are discussed in detail in Appendix B of
Ref.~\cite{ref1}.

The decompositions of $\{\Psi^k\}$ into mode operators have
the same form as in Eqs.~(2.102)--(2.104) of Ref.~\cite{ref2}:
\bea
 \Psi^{k}(z)R_{a}(0) &=&
 \sum_{n}\frac{1}{(z)^{\Delta_{k}+\frac{n}{2}}}\,
 A^{k}_{\frac{n}{2}}R_{a}(0), \qquad k=1,2,\ldots,\frac{N}{2}.
 \label{mode1R}\\
 \Psi^{-k}(z)R_{a}(0) &=& 
\sum_{n}\frac{(-1)^{n}}{(z)^{\Delta_{k}+\frac{n}{2}}}
 A^{1}_{\frac{n}{2}} \, {\sf U}^k R_{a}(0), \qquad
 k=1,2,\ldots,\frac{N}{2} \label{mode2R} \\
 A^{k}_{\frac{n}{2}}R_{a}(0)&=&\frac{1}{4\pi i}\oint_{C_{0}}
 {\rm d}z \, (z)^{\Delta_{k}+\frac{n}{2}-1}\Psi^{k}(z)R_{a}(0),
\label{mode}
\eea
where $a=1,2,\ldots,N$ is the index of the disorder operator, and
${\sf U}$ is a $N \times N$ matrix which rotates this index, ${\sf U}
R_{a}(0) = R_{a-1}(0)$. In Eq.~(\ref{mode}), the action of the
parafermionic modes $ A^{k}_{\frac{n}{2}}$ has been expressed in terms
of contour integrals which are defined by letting $z$ turn twice
around the operator $R_{a}(0)$ at the origin. The representation
(\ref{mode}) has been used to determine the commutation relations
between the modes, given in Eqs.~(2.105)--(2.107) of Ref.~\cite{ref2}.

The decomposition of the local products $\Psi^{k}(z) R_{a}(0)$ into
half-integer powers of $z$ is due to the non-abelian monodromy of the
disorder operator $R_{a}(z,\bar{z})$ with respect to the chiral fields
$\Psi^{\pm k}(z)$, with $k=1,2,\ldots,N/2-1$.

On the other hand, the monodromy of the disorder operator with respect
to the parafermionic field with maximal charge $\Psi^{N/2}$ turns out
to be abelian. Thus, in the expansion of $\Psi^{N/2}(z) R_{a}(0)$ only
integer powers of $z$ are present, as can be seen by
comparing the expansions (\ref{mode1R}) and (\ref{mode2R}), and taking into
account that $\Psi^{-N/2}\equiv\Psi^{N/2}$. In other words, we
have $A^{N/2}_{\frac{n}{2}}R_{a}(0)=0$ for $n$ odd. This peculiarity
marks a difference in the degeneracy structure of the fundamental
disorder operators for the cases of $N$ even and odd.

It is seen from the expansion (\ref{mode1R}) that there are $N/2$
zero modes $A^{k}_{0}$ (with $k=1,2,\ldots,N/2$)
associated with the parafermion $\Psi^k$ which acts between the
$N$ summits of the module:
\beq
 A^{k}_{0}R_{a} = h_{k} \, {\sf U}^{2k}R_{a}. \label{Rzero1}
\eeq
This defines $N/2$ eigenvalues $\{h_{k}\}$ which characterise, in
addition to the conformal dimension $\Delta_R$, each representation
$R_a$. Actually, as described in Ref.~\cite{ref2}, there is a relation
between the eigenvalues $h_1$ and $h_2$ which holds irrespectively of
the details of the representation of a particular operator $R_a$:
\beq
 2h_1^2=\lambda^{1,1}_{2} 2^{\Delta_{2}-3}h_2+ 2^{-\Delta_{2}-2}
 \left[ \kappa(0)+\frac{16\Delta_{1}}{c} \Delta_R \right],
\label{h1h2}
\eeq
where
\beq
 \kappa(n)=
(2\Delta_{1}+n-1)(2\Delta_{1}+n-2)-(2\Delta_{1}+n-1)
 (\Delta_{2}+1)+\frac{(\Delta_{2}+1)(\Delta_{2}+2)}{4}.
\eeq

As witnessed by the expansions (\ref{mode1R})--(\ref{mode2R}),
each disorder module has $N$ summits, labeled by the components
$R_a$, and has only integer and half-integer levels. When  $N$ is
odd, for a given primary operator $R_a$  there are $(N-1)/2$
states at level $1/2$:
\beq
 \left( \chi^{(k)}_{a} \right)_{-\frac{1}{2}} =
 A^{k}_{-\frac{1}{2}} {\sf U}^{-2k}R_{a} =
 {\sf U}^{-2k} A^{k}_{-\frac{1}{2}}R_{a},
 \qquad k=1,2,\ldots,\frac{N-1}{2}. \label{Rstate1}
\eeq
Imposing that all the states (\ref{Rstate1}) be primary, and using
Eq.~(\ref{h1h2}), gives all the $(N+1)/2$ conditions required to
fix the values of all the eigenvalues $\{h_{k}\}$ and of the
conformal dimension $\Delta_R$. We showed in Ref.~\cite{ref2} that the
solutions of the resulting system of equations give the conformal
dimensions of the fundamental disorder operators.

When $N$ is even, we have only $(N-2)/2$ states at level $1/2$:
\beq
 \left( \chi^{(k)}_{a} \right)_{-\frac{1}{2}} =
 A^{k}_{-\frac{1}{2}} {\sf U}^{-2k}R_{a} =
 {\sf U}^{-2k} A^{k}_{-\frac{1}{2}}R_{a},
 \qquad k=1,2,\ldots,\frac{N-2}{2} \label{Rstate2},
\eeq
the ``missing'' state being $A^{N/2}_{-\frac{1}{2}}R_{a}=0$, as we have seen
above. Following the exemple of the case $N$ odd, we shall demand that
all the states (\ref{Rstate2}) be singular. We therefore require that
\beq
 A^{k'}_{+\frac{1}{2}}(\chi^{(k)}_{a})_{-\frac{1}{2}}=0 \label{disdegcond}
\eeq
for each $k'=1,2,\ldots,(N-2)/2$, and for each $k=1,2,\ldots,(N-2)/2$.
The degeneracy condition (\ref{disdegcond}) and the relation (\ref{h1h2})
result in a system of $N/2$ independent equations. The number of
equations is not sufficient to determine the $N/2+1$ unknown
variables, i.e., the $N/2$ zero mode eigenvalues and the conformal
dimension. Therefore---in contradistinction to case of $N$ odd---we shall
have to require the module be degenerate also at the next available
level, i.e., at level 1.

After imposing the condition (\ref{disdegcond}), all the states
(\ref{Rstate2}) can be put equal to zero. After this reduction,
the level $1/2$ will be completely empty. With this
in mind, it is not difficult to verify that at level 1 we
have to consider the state
\beq
 \left( \chi^{(1)}_{a}
 \right)_{-1} =a A^{1}_{-1} {\sf U}^{-2}R_{a} + b L_{-1} R_a
\eeq
and require the following conditions to be satisfied:
\beq
 L_{+1}\left(\chi^{(1)}_{a} \right)_{-1}=0, \quad
 A^{1}_{+1}\left(\chi^{(1)}_{a} \right)_{-1}=0.
\eeq

In terms of the matrix elements $\mu_{ij}$ defined by
\bea
 L_{+1}L_{-1} R_a &=&\mu_{11} R_a, \\
 L_{+1}A_{-1}^{1}{\sf U}^{-2}R_{a} &=& \mu_{12}R_a, \\
 A_{+1}^{-1}L_{-1} R_a &=& \mu_{21}{\sf U}^{2}R_a , \\
 A_{+1}^{-1}A_{-1}^{1}{\sf U}^{-2}R_{a} &=& \mu_{22}{\sf U}^{2}R_a,
\eea
the degeneracy criterion reads
\beq
 \mu_{11}\mu_{22}-\mu_{12}\mu_{21}=0.\label{condlev1}
\eeq
Using the commutation relations given in Eqs.~(2.105)-(2.107)
of Ref.~\cite{ref2}, the matrix elements are readily computed:
\bea
 \mu_{11} &=& 2\Delta_{R}, \\
 \mu_{12} &=& \mu_{21}= \Delta_1 h_{1}, \\
 \mu_{22} &=& \lambda^{1,1}_2 2^{\Delta_2-3}h_{2}+2^{-\Delta_2-2}\kappa(2)
 +2^{-\Delta_2+1}\frac{2\Delta_1}{c}\Delta_R-\gamma h_{1}^2,
\eea
where we used the abbreviation:
\beq
 \gamma=2\Delta_1+4-\frac{3}{2}\Delta_2+\frac{1}{2}\Delta_2^2.
\eeq

By making extensive use of  the commutation relations given in
Ref.~\cite{ref2}, the equations (\ref{h1h2}), (\ref{disdegcond}) and
(\ref{condlev1}) can be solved. They admit two solutions $\Delta_{R}$
for the conformal dimension the fundamental operator $R_a$:
\beq
 \Delta^{(1)}_{R}  = \frac{1}{16}\frac{(N-1)(p+N)}{p}, \qquad
 \Delta^{(2)}_{R} = \frac{1}{16}\frac{(N-1)(p+2-N)}{p+2}. \label{soluR2}
\eeq
The above solutions have the same form as the corresponding
solutions found in Ref.~\cite{ref2} for $N$ odd. They correspond respectively
to:
\beq
 \Delta_{(1,1,\ldots,1,2,1)(1,1,\ldots,1,1,1)}
 \:\:\mbox{and}\:\:\Delta_{(1,1,\ldots,1,1,1)(1,1,\ldots,1,2,1)}
\eeq
with the boundary term
\beq
 B_R=\frac{r-1}{16}.
\eeq

This explicit calculation of the degeneracies of the fundamental
disorder operators confirms the assignment made in Eq.~(\ref{eq22}),
and the value of the boundary term given in Eq.~(\ref{eq29}).
Furthermore, the degeneracy structure is in perfect agreement with the
one predicted by the method of reflections. Following this method it
is easy to see that the module of the operator $R$ possesses $(N-2)/2$
singular states at level $1/2$ (using the $r-1$ simple reflections
$s_a$ with $a=1,2,\ldots,r-2,r$) and one singular state at level $1$
(obtained from the simple reflection $s_{r-1}$).

\end{document}